\documentclass[12pt,superscriptaddress]{revtex4}
\usepackage{hyperref}

\usepackage{amsmath,amsfonts}
\usepackage{amssymb,latexsym}
\usepackage{color}
\usepackage{graphicx}

\newcommand{\ii}{\mathrm{i}}
\newcommand{\ud}{\mathrm{d}}
\newcommand{\um}{\mathrm{m}}
\newcommand{\uD}{\mathrm{D}}
\newcommand{\uM}{\mathrm{M}}
\newcommand{\uE}{\mathrm{E}}
\newcommand{\RE}{\mathrm{R}e}

\begin{document}

\title{Uncertainty Relations in Hydrodynamics}

\author{G.\ Gon\c{c}alves de Matos}
\email{gyell@pq.uenf.br}
\affiliation{Instituto de F\'{\i}sica, Universidade Federal do Rio de Janeiro, C.P. 68528,
21941-972, Rio de Janeiro, RJ, Brazil}
\author{T.\ Kodama}
\email{kodama.takeshi@gmail.com}
\affiliation{Instituto de F\'{\i}sica, Universidade Federal do Rio de Janeiro, C.P. 68528,
21941-972, Rio de Janeiro, RJ, Brazil}
\affiliation{Instituto de F\'{\i}sica, Universidade Federal Fluminense, 24210-346,
Niter\'{o}i, RJ, Brazil}

\author{T.\ Koide}
\email{tomoikoide@gmail.com,koide@if.ufrj.br}
\affiliation{Instituto de F\'{\i}sica, Universidade Federal do Rio de Janeiro, C.P. 68528,
21941-972, Rio de Janeiro, RJ, Brazil}

\begin{abstract}
The uncertainty relations in hydrodynamics are numerically studied. 
We first give a review for the formulation of the generalized uncertainty relations in the stochastic variational method (SVM), following the paper by two of the present authors [Phys.\ Lett.\ A\textbf{382}, 1472 (2018)]. 
In this approach, the origin of the finite minimum value  
of uncertainty is attributed to the non-differentiable (virtual) trajectory of a quantum particle and then 
both of the Kennard and Robertson-Schr\"{o}dinger inequalities in quantum mechanics are reproduced.
The same non-differentiable trajectory is applied to the motion of fluid elements in hydrodynamics.
By introducing the standard deviations of position and momentum for fluid elements, 
the uncertainty relations in hydrodynamics are derived. 
These are applicable even to the Gross-Pitaevskii equation and then the field-theoretical uncertainty relation is reproduced.
We further investigate numerically the derived relations and find that 
the behaviors of the uncertainty relations for liquid and gas are qualitatively different.
This suggests that the uncertainty relations in hydrodynamics are used as a criterion to classify liquid and gas in fluid.
\end{abstract}

\maketitle


\section{Introduction}

The expressions of fundamental laws of physics should be independent of the choice of coordinates. 
This requirement is naturally satisfied when dynamics is formulated in the variational principle.
The variational approach is of wide application and describes the behaviors of particles and fields 
in classical and quantum systems \cite{book:variation}.
On the other hand, there exist several cases where this approach is not applicable.
A dissipative system like hydrodynamics is such an example.
This fact will indicate that the standard formulation of the variational principle can be improved.
As an attempt to describe dissipative systems, Rayleigh's dissipation function method is known \cite{rayleigh}. 
This method is, however, not considered as the improvement of the variational principle, because the dissipative term is introduced by hand as a correction term to the equation obtained by the variational principle.

Note that virtual trajectories considered in the standard variation are implicitly assumed to be smooth 
and non-differentiable trajectories are not considered.
This limitation may be an obstacle to study appropriate optimizations in variation and 
there are several attempts to relax it. 
The generalized framework of the calculus of variations is called stochastic variational method (SVM).
For example, the Navier-Stokes-Fourier (NSF) equation is derived by applying SVM to 
the Lagrangian which leads to the Euler equation in the standard classical variational method 
\cite{inoue,yasue-ns,nakagomi,gomes,eyink,cruzeiro,koide12,del,nov,marner,cruzeiro19}.
Another important aspect is that SVM enables us to describe classical and quantum behaviors in a unified way \cite{yasue}. 
In fact, we can derive the Schr\"{o}dinger equation by applying SVM to the
action which leads to the Newton equation via the classical variation. 
In other words, the quantization of a classical system can be interpreted as the stochastic optimization of a classical action.

Uncertainty relations are known to be an important property in quantum mechanics, which characterizes 
the correlation between fluctuations of, for example, position and momentum of a quantum particle. 
Using this property appropriately \cite{ozawa,ozawa2,ozawa3,ozawa4,ozawa5}, we can investigate a fundamental limitation for simultaneous measurements as was pointed out by Heisenberg \cite{heisenberg}. 
Usually, this property is understood through the non-commutativity of self-adjoint operators. 
If SVM is a natural framework of quantum mechanics, the same property should be obtained 
from the stochasticity of a quantum particle without introducing operators. 
Moreover it will enable us to define the uncertainty relations in hydrodynamics. 
Such a generalized formulation is studied in Refs. \cite{koide18,koide20-1} and  
the Kennard-type and Robertson-Schr\"{o}dinger-type inequalities are derived.

One may wonder that the uncertainty relations in hydrodynamics are hardly surprising because 
those in quantum mechanics can be understood as the wave property of wave function.
It is known that there exists a relation between the fluctuations of position and wave number ($k$) 
when a wave packet is decomposed with plane waves. 
It should be however stressed that this is the relation for position and wave number. 
To obtain the relation associated with momentum, we have to assume the Einstein-de Broglie relation and then $k$ is interpreted as a momentum $\hbar k$.
On the other hand, our uncertainty relations are defined for the canonical momentum 
of fluid elements and are different from the relation for wave number.
Indeed, the minimum value of our relations is affected by viscosity but the corresponding quantity for wave number is irrelevant to dynamics.

In this paper, some properties of the uncertainty relations in hydrodynamics are studied through numerical simulations.  
We first give a review of SVM and the formulation of the uncertainty relations, following Ref.\ \cite{koide18}.
In this approach, the origin of the finite minimum value of uncertainty in quantum mechanics 
is attributed to the ambiguity of the definition of velocity in the non-differentiable (virtual) trajectory of a quantum particle.
After defining canonical momenta in the Hamiltonian formulation of SVM, the uncertainty relations, both of the Kennard and Robertson-Schr\"{o}dinger inequalities, are reproduced.
The same non-differentiable trajectory is applied to the motion of fluid elements in hydrodynamics.
By introducing the standard deviations of position and momentum for fluid elements, 
the Kennard-type and Robertson-Schr\"{o}dinger-type relations are derived for the fluid described by the NSF equation.
These relations are applicable to the trapped Bose gas described by the Gross-Pitaevskii equation and then the field-theoretical uncertainty relation is reproduced. 
After the review, we investigate numerically the derived relations by choosing two parameter set: one is for gas and the other for liquid.
We then find that 
the behaviors of the uncertainty relations for liquid and gas are qualitatively different.
This suggests that the uncertainty relations in hydrodynamics are used as a criterion to classify liquid and gas in fluid.

This paper is organized as follows.
In Sec.\ \ref{sec:clas}, we reformulate the standard variational method in classical particle systems from the point of view of a field theory. This reexpression of the standard classical variation is used as the basis to develop SVM.
Mathematical preparations for stochastic calculus are summarized in Sec. \ref{sec:svm_general} 
and SVM for single-particle systems is discussed in Sec.\ \ref{sec:svm_particle}. 
In Sec.\ \ref{sec:uc_particle}, the Kennard and Robertson-Schr\"{o}dinger inequalities in quantum mechanics are reproduced in SVM. 
The procedure is generalized to continuum media and the uncertainty relations in hydrodynamics are derived in Sec.\ \ref{sec:uc_fluid}. The numerical calculations of the derived relations are shown in Sec.\ \ref{sec:numerical}.
Section \ref{sec:conclusion} is devoted to discussions and concluding remarks.

\section{Classical variation and optimal control} \label{sec:clas}

We briefly summarize the standard variational method in classical mechanics from the perspective of 
an optimal control for a velocity field.

Let us denote the trajectory of a particle ${\bf r}({\bf R},t) = (r^1({\bf R},t)  , \cdots, r^D ({\bf R},t) )$, which is a function of the time $t$ in a spatial $D$-dimensional system of Cartesian coordinates. 
The dependence on the initial position ${\bf r} ({\bf R},t_i)={\bf R}$ at an initial time $t_i$ is explicitly shown. 
The Lagrangian for this single-particle system is defined through the kinetic term $K$ and the potential term $V$ as, 
\begin{eqnarray}
L = K - V =  \frac{\uM}{2} \left( \frac{\ud {\bf r}({\bf R}, t)}{\ud t} \right)^2 - V({\bf r}({\bf R}, t) )\, , \label{eqn:cla-lag}
\end{eqnarray}
where $\ud {\bf r}({\bf R}, t) = {\bf r}({\bf R}, t+\ud t)  - {\bf r}({\bf R}, t) $ and $\uM$ is the particle mass.

For the later convenience, we rewrite this Lagrangian as a field-theoretical quantity.  
Note that the particle probability distribution for this system is defined by 
\begin{eqnarray}
\rho( {\bf x},t ) = \delta^{(D)} ({\bf x}-{\bf r}({\bf R},t) ) \, , \label{eqn:rho_cp}
\end{eqnarray}
where we used Dirac's delta function in $D$ dimension.  
Using this, the above single-particle Lagrangian is given by
\begin{eqnarray}
L = \int \ud^D x \, \rho({\bf x},t) 
\left[
\frac{\uM}{2} {\bf u}^2({\bf x},t) - V({\bf x})
\right] \, ,
\label{eqn:cla-lag-2}
\end{eqnarray}
where ${\bf u}({\bf x},t)$ is a smooth vector function and can be interpreted as the velocity field.
These Lagrangians are equivalent by using the identification 
\begin{eqnarray}
\ud {\bf r}({\bf R}, t) = {\bf u} ({\bf r}({\bf R},t),t) \ud t \, . \label{eqn:cla-vel}
\end{eqnarray}

We consider the time evolution of the particle during the period from an initial time $t_i$ to a final time $t_f$.
For given ${\bf u}({\bf x},t)$ and ${\bf r}({\bf R}, t)$, we can calculate the action defined by 
\begin{eqnarray}
I [{\bf r}] = \int^{t_f}_{t_i} \ud t \, \left[
\frac{\uM}{2} \left( \frac{\ud {\bf r}({\bf R}, t)}{\ud t} \right)^2 - V({\bf r}({\bf R}, t) )
\right] \, .
\end{eqnarray}
The trajectory ${\bf r}({\bf R}, t)$ satisfies Eq.\ (\ref{eqn:cla-vel}) and we determine the form of ${\bf u}({\bf x},t)$ by employing the Hamilton principle.
The infinitesimal variation of trajectories is given by
\begin{eqnarray}
{\bf r}({\bf R}, t) \longrightarrow {\bf r}^\prime ({\bf R}, t) = {\bf r}({\bf R}, t) + \delta {\bf f} ({\bf r}({\bf R}, t),t) \, ,\label{eqn:cla-vari}
\end{eqnarray}
where the infinitesimal smooth function satisfies the boundary conditions $\delta {\bf f}({\bf x},t_i) = \delta {\bf f}({\bf x},t_f) = 0$.
Then the variation of the action leads to
\begin{eqnarray}
\delta I [{\bf r}]
&=& 
\int^{t_f}_{t_i} \ud t \, 
\left[
- \uM (\partial_t +{\bf u}({\bf r}({\bf R}, t),t)\cdot \nabla ){\bf u}({\bf r}({\bf R}, t),t)  - \nabla V({\bf r}({\bf R}, t) )
\right] \cdot \delta {\bf f}  ({\bf r}({\bf R}, t),t) \, .
\label{eqn:deltaI_cla}
\end{eqnarray}
In this calculation, we used
\begin{eqnarray}
{\bf u}^\prime ({\bf r}^\prime({\bf R}, t),t) -  {\bf u}({\bf r}({\bf R}, t),t) 
=
\frac{\ud {\bf r}^\prime ({\bf R}, t)}{\ud t} - \frac{\ud {\bf r} ({\bf R}, t)}{\ud t}
= 
\frac{\ud}{\ud t }\delta {\bf f} ({\bf r}({\bf R}, t),t) \, ,  \label{diff-df}
\end{eqnarray}
where $\ud/\ud t = \partial_t +  \frac{\ud {\bf r}({\bf R}, t)}{\ud t} \cdot \nabla$.
To satisfy $\delta I [{\bf r}]=0$ for an arbitrary function $\delta {\bf f}({\bf x},t)$, the velocity of the particle should satisfy 
the well-known Newton equation, 
\begin{eqnarray}
\uM \frac{\ud}{\ud t} {\bf v}(t) = - \nabla V({\bf r}({\bf R}, t)) \, ,
\end{eqnarray}
where the particle velocity is denoted by
\begin{eqnarray}
{\bf v}(t) = {\bf u}({\bf r}({\bf R}, t),t) \, . 
\end{eqnarray}

\section{General setup for Stochastic variation} \label{sec:svm_general}

In the previous section, 
we implicitly assumed that the particle trajectory is smooth and hence 
the virtual trajectory defined by Eq.\ (\ref{eqn:cla-vari}) is always differentiable. 
As was mentioned in the introduction, there are several proposals to formulate the variation of non-differentiable trajectories 
\cite{inoue,yasue,papiez,yasue-ns,nakagomi,guerra,rosen,marra,serva,kime1989,jae,pav,naga,kappen,gomes,
cre,eyink,cruzeiro,koide12,holm,del,nov,marner,cruzeiro19,kurihara2018,ohsumi2019,lindgren2019,koide19,koide20-2,
holland77,flemin83,ioannis80}.
In this work we use the method proposed by Yasue \cite{yasue} who introduced this idea to extend the formulation of Nelson's stochastic mechanics \cite{nelson}.
Then the variational principle for the classical Lagrangian ($K-V$) 
leads to the Schr\"{o}dinger equation in applying SVM.
See also review papers \cite{zam-review,koide-review2}. As a review of selected topics of stochastic approaches to hydrodynamics, see Ref.\ \cite{cruzeiro2020}.

There are two important aspects in applying the variational principle to the non-differentiable trajectory: 
one is the introduction of two Brownian motions (see Eqs.\ (\ref{eqn:fsde}) and (\ref{eqn:bsde})) 
and the other consists in the generalization of the time derivatives (see Eqs.\ (\ref{eqn:mfd}) and (\ref{eqn:mbd})).

\subsection{Zigzag trajectory and two Brownian motions}

\begin{figure}[t]
\begin{center}
\includegraphics[scale=0.4]{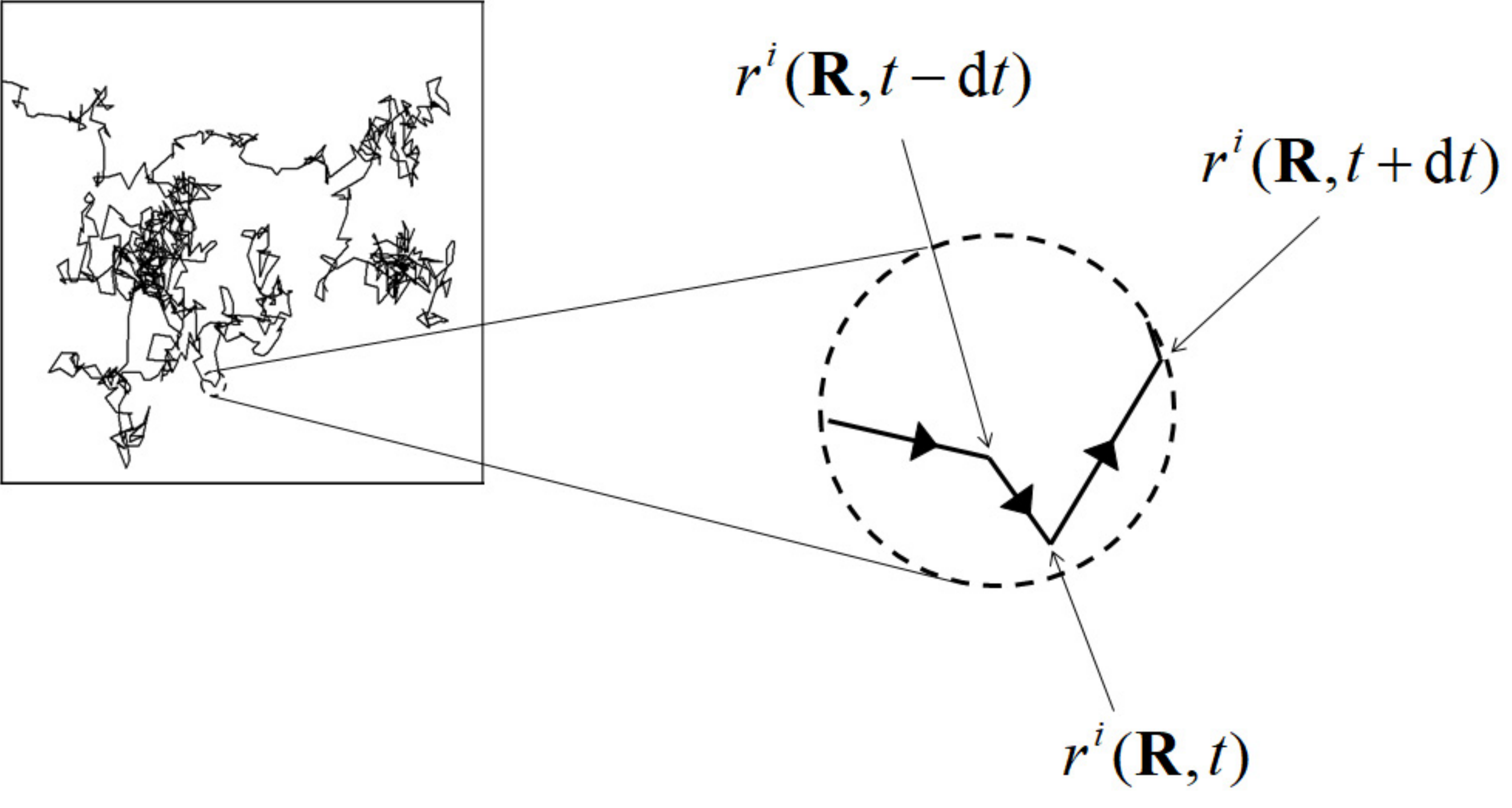}
\end{center}
\caption{An example of the typical trajectory of Brownian motion. The positions at $t- \ud t$, $t$ and $t+ \ud t$ are denoted by $r^{i}_{t-\ud t}$, $r^{i}_{t}$ and $r^{i}_{t+\ud t}$, respectively.}
\label{fig:tra1}
\end{figure}

As was discussed in Sec.\ \ref{sec:clas}, 
a particle velocity ${\bf u}({\bf r}({\bf R}, t),t)$ is the tangential line of 
a particle trajectory ${\bf r}({\bf R}, t)$ and satisfies Eq.\ (\ref{eqn:cla-vel}).
Let us generalize this relation to formulate the variation of stochastic variables.

A typical example of a zigzag trajectory is known in Brownian motion as is shown in Fig.\ \ref{fig:tra1}. 
We assume the similar behavior to represent a zigzag trajectory in SVM.  
The simplest way to realize this idea is to add a noise term of Brownian motion to Eq.\ (\ref{eqn:cla-vel}). 
The derived equation describes the forward time evolution of the trajectory and called the forward stochastic differential equation (SDE), 
\begin{eqnarray}
\ud \widehat{\bf r}({\bf R}, t) =  {\bf u}_+ (\widehat{\bf r}({\bf R}, t), t) \ud t + \sqrt{2\nu} \ud\widehat{\bf W}(t)  \, \, \, (\ud t > 0) \, . \label{eqn:fsde}
\end{eqnarray} 
The second term on the right-hand side represents the noise term of Brownian motion and hence the origin of the zigzag motion. 
We used $(\widehat{ \, \, \, \, })$ to denote stochastic variables 
and $\ud \widehat{A}(t) \equiv \widehat{A}(t+ \ud t) - \widehat{A}(t)$ for an arbitrary 
stochastic quantity $\widehat{A}(t)$.
The time variable is discretized with the time width $\ud t$ but we should 
take the vanishing limit of $\ud t$ at the last of calculations.
Here $\widehat{\bf W}(t)$ describes the standard Wiener process which satisfies the following correlation properties, 
\begin{equation}
\begin{split}
\uE[ \ud\widehat{\bf W}(t) ] &= 0\, , \\
\uE [ \ud\widehat{W}^{i} (t) \ud\widehat{W}^{j} (t^\prime) ] &= \ud t \, \delta_{t,t^\prime} \, \delta_{i,j}   \, \, \, (i\,j = x,y,z) \, , 
\end{split}
\label{eqn:wiener}
\end{equation}
where $\uE[\, \, \, \,]$ denotes the ensemble average for the Wiener process and the right-hand side of the second equation is expressed by Kronecker's delta functions.
The  intensity of fluctuations is controlled by a non-negative real constant $\nu$. 
For the detailed description of the Wiener process, see, for example, Ref. \cite{book:gardiner}.
Note that ${\bf u}_+ (\widehat{\bf r}({\bf R}, t),t)$ is stochastic because of 
the trajectory $\widehat{\bf r}({\bf R}, t)$, but ${\bf u}_+ ({\bf x},t)$ is a smooth function.

The particle velocity (field) is, however, not completely characterized by ${\bf u}_+ ({\bf x},t)$ alone. 
As is seen from Fig.\ \ref{fig:tra1}, there are at least two different possibilities to characterize the particle velocities at the particle position $\widehat{\bf r}({\bf R}, t)$: 
one is ${\displaystyle \lim_{\ud t \rightarrow 0+}}(\widehat{\bf r}({\bf R}, t+\ud t) - \widehat{\bf r}({\bf R}, t))/\ud t$ and 
the other ${\displaystyle \lim_{\ud t \rightarrow 0+}}(\widehat{\bf r}({\bf R}, t) - \widehat{\bf r}({\bf R}, t-\ud t))/\ud t$.
These two quantities are equivalent in a smooth trajectory, but different in a zigzag trajectory.
This difference plays an important role in SVM.

The forward SDE is associated with the former definition, ${\displaystyle \lim_{\ud t \rightarrow 0+}}(\widehat{\bf r}({\bf R}, t+\ud t) - \widehat{\bf r}({\bf R}, t))/\ud t$. 
To accommodate the latter definition,  
we introduce the backward time evolution of the trajectory described by the backward SDE, 
\begin{eqnarray}
\ud \widehat{\bf r}({\bf R}, t) =  {\bf u}_- (\widehat{\bf r}({\bf R}, t), t) \ud t + \sqrt{2\nu} \ud \underline{\widehat{\bf W}}(t)  \, \, \, (\ud t < 0) \, , \label{eqn:bsde}
\end{eqnarray} 
where the another standard Wiener process $\underline{\widehat{\bf W}}(t)$ satisfies 
\begin{equation}
\begin{split}
\uE[ \ud\underline{\widehat{\bf W}}(t) ] &= 0\, , \\
\uE [ \ud\underline{\widehat{W}}^{i} (t) \ud\underline{\widehat{W}}^{j} (t^\prime) ] &= |\ud t| \, \delta_{t,t^\prime}\, \delta_{i,j}   \, \, \, (i\,j = x,y,z) \, .
\end{split}
\label{eqn:wiener2}
\end{equation}
These are the same as those in Eq.\ (\ref{eqn:wiener}) by replacing $\ud t$ with $|\ud t|$.

\subsection{Mean forward and backward derivatives}

\begin{figure}[t]
\begin{center}
\includegraphics[scale=0.3]{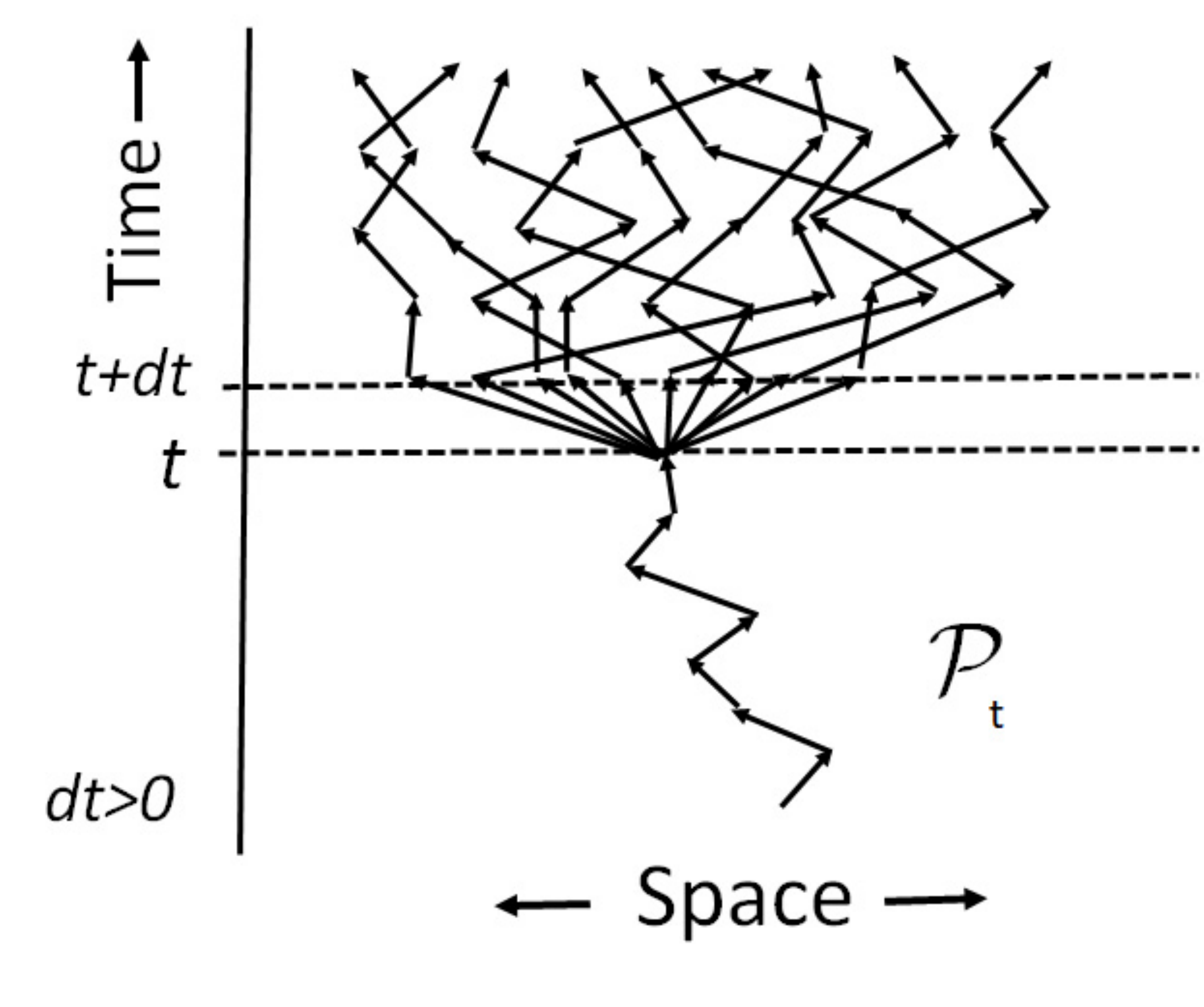}
\includegraphics[scale=0.3]{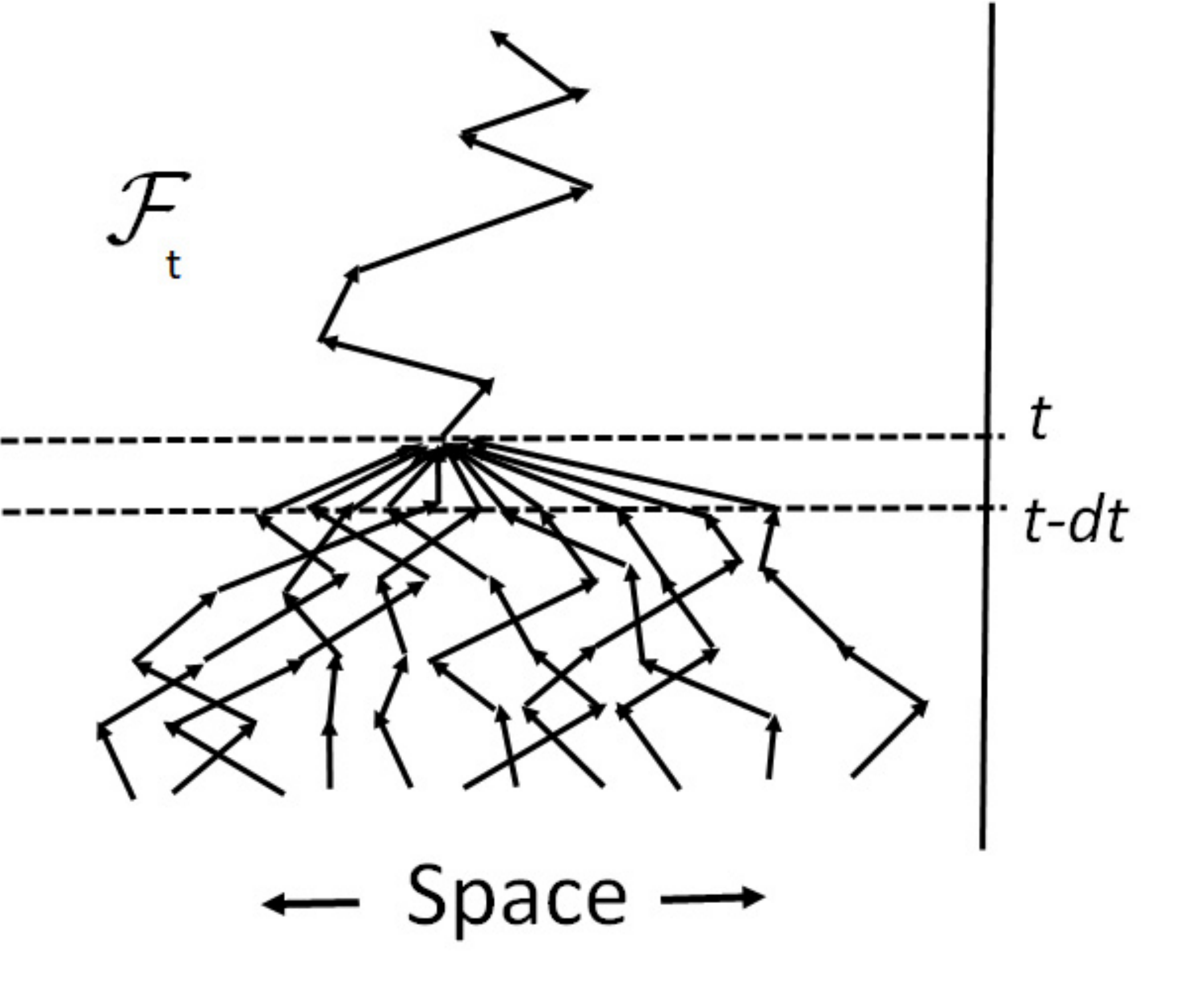}
\end{center}
\caption{The ensembles of trajectories fixing ${\cal P}_t$ and ${\cal F}_t$ are represented on 
the left and right panels, respectively.  }
\label{fig:tra2}
\end{figure}

Because of the ambiguity of the velocity mentioned above, 
Nelson introduced two different time derivatives in stochastic mechanics \cite{nelson};
 one is the mean forward derivative,
\begin{equation}
\uD_+  f  ( \widehat{\bf r}({\bf R}, t) )  = \lim_{\ud t \rightarrow0+} \uE \left[  \frac{ f(\widehat{\bf r}({\bf R}, {t + \ud t})) -
f(\widehat{\bf r}({\bf R}, t))}{\ud t} \Big| \mathcal{P}_{t} \right]\,  , \label{eqn:mfd}
\end{equation}
and the other the mean backward derivative,
\begin{equation}
\uD_-  f(\widehat{\bf r}({\bf R}, t))  = \lim_{\ud t \rightarrow0-} \uE \left[  \frac{ f(\widehat{\bf r}({\bf R}, {t + \ud t})) -
f(\widehat{\bf r}({\bf R}, t))}{\ud t} \Big| \mathcal{F}_{t} \right]\, . \label{eqn:mbd}
\end{equation}
These expectation values are conditional averages, where $\mathcal{P}_{t}$ ($\mathcal{F}_{t}$) indicates  
a set of trajectories fixing $\widehat{\bf r}_{t^\prime}$ for
$t^{\prime}\le t~~(t^{\prime}\ge t)$. 
As a schematic figure for these definitions, see Fig.\ \ref{fig:tra2}.
For the $\sigma$-algebra of all measurable events of $\widehat{\bf r}_{t}$, $\mathcal{P}_{t}$ and $\mathcal{F}_{t}$
represent an increasing and a decreasing family of sub-$\sigma$-algebras, respectively.

To elucidate the above definitions, we consider the Markov process where the trajectory at $t+ \ud t$ 
is determined from the position at $t$ and independent of the hysteresis of the position  
${\bf r}({\bf R}, t^\prime)$ $(t ^\prime < t)$ as is described by the forward SDE.
Then the condition denoted by $\mathcal{P}_{t}$ is simplified so that only the position at $t$ is fixed and then 
the above definition of the mean forward derivative becomes 
\begin{equation}
\uD_+  f(\widehat{\bf r}({\bf R}, t))  = \lim_{\ud t \rightarrow0+} \uE \left[  \frac{ f(\widehat{\bf r}({\bf R}, {t + \ud t})) -
f(\widehat{\bf r}({\bf R}, t))}{\ud t} \Big| \widehat{\bf r}({\bf R}, t) \right]\, .
\end{equation}
To understand this, let us consider a class of the stochastic trajectories which passes a fixed position 
$\widehat{\bf r} ({\bf R}, t)$ at $t$. 
As is shown on the left panel of Fig.\ \ref{fig:tra2}, the position of the particle at $t+\ud t$ still fluctuates and then we can consider the average for such a fluctuating position.
Substituting the forward SDE into the mean forward derivative, we easily find  
\begin{eqnarray}
\uD_+  \widehat{\bf r}({\bf R}, t)  = \lim_{\ud t \rightarrow0+} \uE \left[  {\bf u}_+ (\widehat{\bf r}({\bf R}, t),t) + \sqrt{2\nu} \frac{\ud {\bf W}(t)}{\ud t} \Big| \widehat{\bf r}({\bf R}, t) \right] =  {\bf u}_+ (\widehat{\bf r}({\bf R}, t),t)\, .
\end{eqnarray}
In the second equality, we used the first equation of Eq.\ (\ref{eqn:wiener}).
In a similar fashion, the average for the position $\widehat{\bf r}(t-\ud t)$ is considered for the backward SDE, 
as is shown on the right panel of  Fig.\ \ref{fig:tra2} 
and then 
the mean backward derivative (\ref{eqn:mbd}) leads to 
\begin{eqnarray}
\uD_-  \widehat{\bf r}({\bf R}, t)  = \lim_{\ud t \rightarrow0-} \uE \left[  {\bf u}_- (\widehat{\bf r}({\bf R}, t),t) + \sqrt{2\nu} \frac{\ud \underline{\bf W}(t)}{\ud t} \Big| \widehat{\bf r}({\bf R}, t) \right] =  {\bf u}_- (\widehat{\bf r}({\bf R}, t),t)\, .
\end{eqnarray}

These derivatives are related through the stochastic partial integration formula,
\begin{eqnarray}
\lefteqn{ \int^{t_f}_{t_i} \ud t  \, 
\uE\left[
\widehat{Y}(t) \uD_+ \widehat{X}(t) + \widehat{X}(t) \uD_- \widehat{Y}(t)
\right] } && \nonumber \\
&=& 
\lim_{n \rightarrow \infty} \sum_{j=0}^{n-1} \uE\left[
(\widehat{X}(t_{j+1}) - \widehat{X}(t_j)) \frac{\widehat{Y}(t_{j+1}) + \widehat{Y}(t_j)}{2}
\right] 
 + 
\lim_{n \rightarrow \infty} \sum_{j=1}^{n} \uE\left[
(\widehat{Y}(t_{j}) - \widehat{Y}(t_{j-1})) \frac{ \widehat{X}(t_j) + \widehat{X}(t_{j-1})}{2}
\right] \nonumber \\
&=& 
\lim_{n \rightarrow \infty} \sum_{j=0}^{n-1} \uE\left[
\widehat{X}(t_{j+1}) \widehat{Y}(t_{j+1})-\widehat{X}(t_{j})\widehat{Y}(t_{j})
\right]  \nonumber \\
&=& 
\uE\left[
\widehat{X}(t_f)\widehat{Y}(t_f) - \widehat{X}(t_i)\widehat{Y}(t_i)
\right] \, , \label{eqn:spif}
\end{eqnarray}
where $t_j = t_i + j \frac{t_f - t_i}{n} = t_i + j \ud t$.

\subsection{Fokker-Planck equation and Consistency condition} \label{sec:fp}

The two SDE's should describe the same stochastic process.
To satisfy this requirement, the two velocity fields ${\bf u}_\pm ({\bf x},t)$ cannot be independent.

To see this, we define the particle probability distribution 
\begin{eqnarray}
\rho ({\bf x},t) = \int \ud^D R \,  \rho_0 ({\bf R}) \uE\left[
\delta^{(D)} ({\bf x} - \widehat{\bf r}({\bf R}, t))
\right]   \, , \label{eqn:rho}
\end{eqnarray}
where 
$\rho_0 ({\bf R})$ is the distribution of the initial position of the particle satisfying 
\begin{eqnarray}
\int \ud^D R \, \rho_0 ({\bf R}) = 1 \, .
\end{eqnarray}

As is well-known, the evolution equation of $\rho ({\bf x},t)$ is obtained by using the SDE's and Ito's lemma \cite{book:gardiner} which is a truncated Taylor series expansion for a function with stochastic variables 
as is summarized in Appendix\ \ref{app:ito}. 
The derived equation is known as the Fokker-Planck equation. By using the forward SDE in the derivation, this  becomes 
\begin{eqnarray}
\partial_t \rho ({\bf x},t) = ( - \nabla \cdot {\bf u}_+ ({\bf x},t ) + \nu \nabla^2 ) \rho ({\bf x},t)  \, .
\label{eqn:ffp}
\end{eqnarray}
Another Fokker-Planck equation from the backward SDE is given by
\begin{eqnarray}
\partial_t \rho ({\bf x},t) = ( - \nabla \cdot {\bf u}_- ({\bf x},t ) - \nu \nabla^2 ) \rho ({\bf x},t)  \, .
\label{eqn:bfp}
\end{eqnarray}
The forward and backward SDE's describe the different aspect of the same stochastic trajectory and hence the above two Fokker-Planck equations must be equivalent. 
Therefore the following condition should be satisfied
\begin{eqnarray}
 {\bf u}_+ ({\bf x},t) =  {\bf u}_- ({\bf x},t) + 2\nu \nabla  \ln \rho ({\bf x},t) 
+ \frac{1}{\rho ({\bf x},t)} \nabla \times {\bf A} ({\bf x}, t)
\, ,
\end{eqnarray}
where  ${\bf A} ({\bf x},t)$ is an arbitrary vector function. 
For the sake of simplicity, we set ${\bf A} ({\bf x},t) = 0$ and then the above equation is simplified as 
\begin{eqnarray}
 {\bf u}_+ ({\bf x},t) =  {\bf u}_- ({\bf x},t) + 2\nu \nabla \ln \rho ({\bf x},t) \label{eqn:cc}
\, .
\end{eqnarray}
This is called the consistency condition.
The possible role of the omitted function ${\bf A}({\bf x},t)$ is not yet understood.

We are often interested in the velocity field ${\bf v} ({\bf x},t)$ which is parallel to the current of the particle probability distribution,
\begin{eqnarray}
\rho ({\bf x},t) {\bf v} ({\bf x},t)
= \rho ({\bf x},t){\bf u}_+ ({\bf x},t) - \nu \nabla \rho ({\bf x},t) 
= \rho ({\bf x},t) {\bf u}_- ({\bf x},t) + \nu \nabla \rho ({\bf x},t) \, .
\end{eqnarray}
By using the consistency condition, we find that ${\bf v} ({\bf x},t)$ is given by the average of ${\bf u}_\pm ({\bf x},t)$,
\begin{eqnarray}
{\bf v} ({\bf x},t) = \frac{{\bf u}_+ ({\bf x},t) + {\bf u}_- ({\bf x},t)}{2} \, . \label{eqn:ave-vel}
\end{eqnarray}
Using this velocity field, the two Fokker-Planck equations are reduced to 
the simple equation of continuity, 
\begin{equation}
\begin{split}
\partial_t \rho ({\bf x},t) &= - \nabla \cdot {\bf J}({\bf x},t) \, ,\\
{\bf J} ({\bf x},t) &= \rho({\bf x},t) {\bf v}({\bf x},t) \, .
\end{split}
\label{eqn:eoc}
\end{equation}

The consistency condition is a key property to obtain the uncertainty relations in this formulation.
In fact, multiplying $x^i$ to the condition, the expectation value is calculated as  
\begin{eqnarray}
\int \ud^D x \, \rho ({\bf x},t) \left\{ x^i u^j_- ({\bf x},t) - x^i u^j_+ ({\bf x},t) \right\} = 2\nu \delta_{ij} \, .
\end{eqnarray}
As is seen later, we set $\nu = \hbar/(2\uM)$ to reproduce the Schr\"{o}dinger equation in the SVM quantization. Then this equation reminds us the expectation value of the canonical commutation rule.
See also Ref.\ \cite{biane2010}.

\section{Stochastic variational method for particle} \label{sec:svm_particle}

We consider the stochastic variation for single-particle systems where
the classical form of the Lagrangian is given by Eq.\ (\ref{eqn:cla-lag}).

\subsection{Stochastic action and its variation}

The corresponding Lagrangian for the stochastic variation is obtained by replacing ${\bf r}({\bf R}, t)$ with 
$\widehat{\bf r}({\bf R}, t)$. 
Such a replacement is however not trivial for the kinetic term 
because there are now two possibilities to replace the time derivative of smooth trajectories. 
In this work, we assume that the stochastic representation of the kinetic term is given by the most general
quadratic form of the two derivatives,
\begin{eqnarray}
\frac{\uM}{2} \left[
A  (\uD_+ \widehat{\bf r}({\bf R}, t) )^2 + B  (\uD_- \widehat{\bf r}({\bf R}, t) )^2 
+ C (\uD_+ \widehat{\bf r}({\bf R}, t) ) \cdot (\uD_- \widehat{\bf r}({\bf R}, t) )
\right] \, ,
\end{eqnarray} 
where $A$, $B$ and $C$ are real constants. 
We require that SVM reproduces the result of the classical variation in the vanishing limit of the stochasticity. 
In this limit, both of $\uD_+ \widehat{\bf r}({\bf R}, t)$ and $\uD_- \widehat{\bf r}({\bf R}, t)$ coincide with the standard time derivative of a smooth trajectory. 
Thus, to reproduce the result of the classical variation in the limit, the three coefficients should satisfy 
\begin{eqnarray}
A + B + C = 1 \, .
\end{eqnarray}
Therefore the coefficients can be parametrized as \cite{koide-review1},
\begin{eqnarray}
\frac{\uM}{2} \dot{\bf r}^2({\bf R}, t)  
\rightarrow 
\frac{\uM}{2} 
\left[
B_+ \sum_{l=\pm} A_l (\uD_l \widehat{\bf r}({\bf R}, t) )^2 + B_- (\uD_+ \widehat{\bf r}({\bf R}, t) ) \cdot (\uD_- \widehat{\bf r}({\bf R}, t) )
\right] \, , \label{eqn:quad_rep}
\end{eqnarray}
where 
\begin{equation}
\begin{split}
A_\pm &= \frac{1}{2} \pm \alpha_A \, ,\\
B_\pm &= \frac{1}{2} \pm \alpha_B \, ,
\end{split}
\end{equation}
with $\alpha_A$ and $\alpha_B$ being real constants. 
The right-hand side of 
Eq.\ (\ref{eqn:quad_rep}) coincides with the left-hand side in the vanishing limit of $\nu$ because
the two time derivatives are reduced to the time derivative of a smooth trajectory.

In the classical variation, the form of a trajectory is entirely controlled for a given velocity. 
In SVM, however, we cannot determine a trajectory without ambiguity even for a given velocity field due to the stochasticity 
of the trajectory. 
Therefore we should consider the optimization of the averaged behavior of an action.
The action is eventually given by the expectation value, 
\begin{eqnarray}
I_{sto} [\widehat{\bf r}] 
= \int^{t_f}_{t_i} \ud t \, \uE[L_{sto}(\widehat{\bf r},\uD_+ \widehat{\bf r}, \uD_- \widehat{\bf r})]\, ,
\label{eqn:sto_act}
\end{eqnarray}
where the stochastic Lagrangian is defined by
\begin{eqnarray}
L_{sto} (\widehat{\bf r},\uD_+ \widehat{\bf r}, \uD_- \widehat{\bf r}) 
= 
\frac{\uM}{2} (\uD_+ \widehat{\bf r}({\bf R}, t), \uD_- \widehat{\bf r}({\bf R}, t)) 
{\cal M} 
\left(
\begin{array}{c}
\uD_+ \widehat{\bf r}({\bf R}, t) \\
\uD_- \widehat{\bf r}({\bf R}, t)
\end{array}
\right) - V(\widehat{\bf r}({\bf R}, t))
\, , \label{eqn:sto-lag}
\end{eqnarray}
with
\begin{eqnarray}
{\cal M} = 
\left(
\begin{array}{cc}
A_+ B_+ & \frac{1}{2} B_- \\
\frac{1}{2} B_-  & A_- B_+ 
\end{array}
\right) \, . \label{eqn:cal_m}
\end{eqnarray}

For the variation of the stochastic trajectory, instead of Eq.\ (\ref{eqn:cla-vari}), we consider 
\begin{eqnarray}
\widehat{\bf r}({\bf R}, t) \longrightarrow \widehat{\bf r}^\prime ({\bf R}, t) = \widehat{\bf r}({\bf R}, t) + \delta {\bf f} (\widehat{\bf r}({\bf R}, t),t) \, .\label{eqn:csto-vari}
\end{eqnarray}
The infinitesimal smooth function $\delta {\bf f} ({\bf x},t)$ satisfies the same properties defined in Sec. \ref{sec:clas}, 
$\delta {\bf f}({\bf x},t_i) = \delta {\bf f}({\bf x},t_f) = 0$. 
Note that the fluctuation of $\delta {\bf f} (\widehat{\bf r}({\bf R}, t),t) $ comes from that of $\widehat{\bf r}({\bf R}, t)$.

The important feature of the stochastic variation comes from the kinetic term 
which has the time derivatives.
For example, the stochastic variation of a part of the kinetic term is calculated as 
\begin{eqnarray}
\lefteqn{ \int^{t_f}_{t_i} \ud t \,
\uE\left[
(\uD_+ \widehat{\bf r}^\prime ({\bf R}, t) )^2 - (\uD_+  \widehat{\bf r} ({\bf R}, t) )^2 
\right] } && \nonumber \\
&=&
2  \int^{t_f}_{t_i} \ud t \,
\uE\left[ 
{\bf u}_+ (\widehat{\bf r} ({\bf R}, t),t ) \cdot \uD_+ \delta {\bf f} (\widehat{\bf r}({\bf R}, t),t)
\right] \nonumber \\
&=&
- 2  \int^{t_f}_{t_i} \ud t \,
\uE\left[ 
\uD_- {\bf u}_+ (\widehat{\bf r} ({\bf R}, t),t ) \cdot  \delta {\bf f} (\widehat{\bf r}({\bf R}, t),t)
\right] \nonumber \\
&=&
- 2  \int^{t_f}_{t_i} \ud t \,
\uE\left[ 
(\partial_t + {\bf u}_- (\widehat{\bf r}({\bf R}, t),t) \cdot \nabla - \nu \nabla^2)  {\bf u}_+ (\widehat{\bf r} ({\bf R}, t),t ) \cdot  \delta {\bf f} (\widehat{\bf r}({\bf R}, t),t)
\right] \, . 
\end{eqnarray}
From the second to the third line, we used the stochastic partial integration formula (\ref{eqn:spif}).
In the last line, Ito's lemma is utilized.

On the other hand, the variation of the terms independent of time derivatives is the same as 
that in the corresponding classical variation.
Summarizing these results, the variation of the stochastic action is calculated as
\begin{eqnarray}
\delta I [\widehat{\bf r}]
=
 \int^{t_f}_{t_i} \ud t \, 
\uE\left[
{\bf I}_\delta (\widehat{\bf r}({\bf R},t),t)
 \cdot \delta {\bf f} (\widehat{\bf r}({\bf R}, t),t)
\right] 
\label{eqn:deltai1}
\, ,
\end{eqnarray}
where
\begin{eqnarray}
\lefteqn{{\bf I}_\delta ({\bf x},t) } && \nonumber \\
&=& 
-\uM B_+ \left\{ A_+ (\partial_t + {\bf u}_- ({\bf x},t) \cdot \nabla - \nu \nabla^2)  {\bf u}_+ ({\bf x} ,t ) 
 + A_-  (\partial_t + {\bf u}_+ ({\bf x} ,t ) \cdot \nabla + \nu \nabla^2)  {\bf u}_- ({\bf x} ,t ) \right\}
\nonumber \\
&& -
\uM \frac{B_- }{2} 
\left\{  (\partial_t + {\bf u}_- ({\bf x} ,t ) \cdot \nabla - \nu \nabla^2)  {\bf u}_- ({\bf x} ,t ) 
+ 
 (\partial_t + {\bf u}_+ ({\bf x} ,t ) \cdot \nabla + \nu \nabla^2)  {\bf u}_+ ({\bf x} ,t )  \right\} 
\nonumber \\
&& - \nabla V ({\bf x}) \, .
\end{eqnarray}
From the perspective of the classical variational approach, 
one may consider that the vanishing variation of the action $\delta I [\widehat{\bf r}]= 0$ is satisfied 
for any choice of $\delta {\bf f} (\widehat{\bf r}({\bf R},t),t)$ and find ${\bf I}_\delta (\widehat{\bf r}({\bf R},t),t) =0$.
However, we can control only the averaged behavior as was discussed above.
Note that Eq.\ (\ref{eqn:deltai1}) is reexpressed as 
\begin{eqnarray}
\int \ud^D R \, \rho_0 ({\bf R})\delta I [\widehat{\bf r}]
=
  \int^{t_f}_{t_i} \ud t \int \ud^D x \, \rho({\bf x},t)\, 
{\bf I}_\delta ({\bf x},t)
 \cdot \delta {\bf f} ({\bf x},t) =0
\label{eqn:deltai2}
\, ,
\end{eqnarray}
where we multiply the arbitrary distribution of the initial position $\rho_0 ({\bf R})$ to introduce $\rho ({\bf x},t)$ defined by Eq.\ (\ref{eqn:rho}).
Thus the stochastic Hamilton principle requires not ${\bf I}_\delta (\widehat{\bf r}({\bf R},t),t) =0$ 
but ${\bf I}_\delta ({\bf x},t) =0$,  
\begin{eqnarray}
\left(\partial_t + {\bf u}_- ({\bf x},t) \cdot \nabla - \nu \nabla^2, \,  \partial_t + {\bf u}_+ ({\bf x},t) \cdot \nabla + \nu \nabla^2 \right)
{\cal M} 
\left(
\begin{array}{c}
{\bf u}_+ ({\bf x},t ) \\
{\bf u}_- ({\bf x},t ) 
\end{array}
\right)
= - \frac{1}{\uM}\nabla V ({\bf x}) \, . \label{eqn:vari-particle1}
\end{eqnarray}
This equation reproduces the Newton equation in the vanishing limit of fluctuations, 
$\nu \rightarrow 0$. See also Eq.\ (\ref{eqn:deltaI_cla}).
That is, SVM is the natural generalization of the classical variational method.

It is worth mentioning that the above result of the stochastic variation is formally represented by 
\begin{eqnarray}
\left[ \uD_+ \frac{\partial L_{sto}}{\partial (\uD_- \widehat{\bf r})} + \uD_- \frac{\partial L_{sto}}{\partial (\uD_+ \widehat{\bf r})} - \frac{\partial L_{sto}}{\partial \, \widehat{\bf r}}
\right]_{\widehat{\bf r}({\bf R},t) = {\bf x}} = 0 \, .
\label{eqn:sel}
\end{eqnarray}
This is the stochastic Euler-Lagrange equation. 
For the special case where $(\alpha_A,\alpha_B) = (0,1/2)$, 
the stochastic Euler-Lagrange equation coincides with the equation obtained by Nelson in Ref.\ \cite{nelson}.

\subsection{Schr\"{o}dinger equation}

As one of the SVM applications, we consider the quantization based on SVM \cite{yasue}.
For this, we choose $(\alpha_A, \alpha_B) = (0, 1/2)$ which gives ${\cal M} = {\rm diag} (1/2,1/2)$.
Then Eq. (\ref{eqn:vari-particle1}) is simplified as 
\begin{eqnarray}
 (\partial_t + {\bf v}({\bf x},t) \cdot \nabla) {\bf v}({\bf x},t) = - \frac{1}{\uM} \nabla ( V({\bf x}) + V_Q ({\bf x},t) ) \, ,
\label{eqn:vari_qh}
\end{eqnarray}
where the second term on the right-hand side is known as the gradient of the quantum potential defined by 
\begin{eqnarray}
V_Q ({\bf x},t)= -  2 \uM \nu^2 \rho^{-1/2}({\bf x},t) 
\nabla^2 \sqrt{\rho ({\bf x},t)} \, .
\end{eqnarray}
To obtain this expression, we used the averaged velocity (\ref{eqn:ave-vel}) and the consistency condition (\ref{eqn:cc}) to eliminate ${\bf u}_\pm ({\bf x},t)$.
The evolution of $\rho({\bf x},t)$ is given by the equation of continuity (\ref{eqn:eoc}).
That is, the SVM optimization for the single-particle Lagrangian leads to the coupled equations of Eq.\ (\ref{eqn:eoc}) and Eq.\ (\ref{eqn:vari_qh}).

The coupled equations can be represented in a familiar form. 
Let us introduce a scalar function $\theta ({\bf x},t)$, which corresponds to the velocity potential, as
\begin{eqnarray}
{\bf v}({\bf x},t) = 2\nu \nabla \theta ({\bf x},t) \, . \label{eqn:phase}
\end{eqnarray}
This procedure is applicable when the motion is irrotational. 
Then the equation obtained by the stochastic variation (\ref{eqn:vari-particle1}) 
is reexpressed as
\begin{eqnarray}
\partial_t \theta ({\bf x},t) + \nu (\nabla \theta ({\bf x},t) )^2 = - \frac{1}{2 \nu \uM} V ({\bf x}) + \nu\rho^{-1/2}({\bf x},t) \nabla^2 \sqrt{\rho ({\bf x},t)} \, .
\end{eqnarray}
We further introduce a complex function defined by 
\begin{eqnarray}
\Psi ({\bf x},t) = \sqrt{\rho ({\bf x},t)} e^{\ii \theta ({\bf x},t)} \, , \label{eqn:wf}
\end{eqnarray}
which automatically satisfies 
\begin{eqnarray}
|\Psi ({\bf x},t)|^2 = \rho ({\bf x},t) \, .
\end{eqnarray}
The equation of $\Psi ({\bf x},t)$ is easily found from the evolution equations for 
$\rho({\bf x},t)$ and $\theta({\bf x},t)$,
\begin{eqnarray}
\ii \partial_t \Psi ({\bf x},t) = \left[ -\nu \nabla^2 + \frac{1}{2\nu \uM} V({\bf x}) \right]\Psi ({\bf x},t) \, .
\end{eqnarray}
This equation is the Schr\"{o}dinger equation by choosing 
\begin{eqnarray}
\nu = \frac{\hbar}{2\uM} \, ,
\end{eqnarray}
and then $\Psi ({\bf x},t)$ is identified with the wave function.
In fact, Eq. (\ref{eqn:vari_qh}) is known as the hydrodynamical representation of the Schr\"{o}dinger equation 
which was firstly proposed by Madelung \cite{madelung} and has been studied exclusively by Bohm and his collaborators 
\cite{bohm,book:holland,sanz}.
See also the Euler equation of the ideal fluid shown by Eq.\ (\ref{eqn:euler_eq}) for comparison.
From this quantum hydrodynamical perspective, 
quantum effects described by the Schr\"{o}dinger equation are induced by the quantum potential 
$V_Q ({\bf x},t)$.

The Bernoulli equation is known to characterize a conservation law of the ideal fluid. 
Applying the Bernoulli equation to quantum hydrodynamics (\ref{eqn:vari_qh}), 
the time-independent Schr\"{o}dinger equation is obtained. See Appendix \ref{app:ber}.

From the perspective of SVM, 
the quantization of classical systems is interpreted as the variation of the action with more microscopic precision.
Suppose that a microscopically non-differentiable trajectory seems to be smooth 
in a coarse-grained scale. 
When we are interested in macroscopic behaviors where the non-differentiability is negligibly small, 
the classical variation is applicable to an action and the Newton equation is obtained.
On the other hand, in the observation with microscopic scales, 
SVM should be applied and the Schr\"{o}dinger equation is obtained.

In Ref.\ \cite{nelson2}, 
Nelson discussed that the particle introduced for the variation in SVM seems to permit the instantaneous transmission of signals between two distant systems.
Indeed, it is known that the diffusion type-equation like Eqs.\ (\ref{eqn:ffp}) and (\ref{eqn:bfp}) exhibits a superluminal propagation. See, for example, Ref.\ \cite{koide-diff} and references therein.  
We would like to, however, emphasize that the particle considered for the variation is not necessarily a real object,
but rather a mathematical notion to help the formulation of the stochastic variation. 
This view is more apparent in the application to fluids where the motion of the fluid element is considered.

In quantum mechanics, the wave function is required to be a continuous single-valued function.
The corresponding property is however not considered in quantum hydrodynamics.
As a matter of fact, the quantum potential becomes singular at the nodes of wave functions 
and hence we need additional condition to connect the solutions for the left and right sides of the singularity.
The standard procedure is to use the Bohr-Sommerfeld type condition for a loop path of the quantum fluid. 
See Ref.\ \cite{book:holland,koide20-1,takabayasi,wallstrom} for details. 
It is worth mentioning that quantum hydrodynamics has an advantage to discuss quantum behaviors in generalized coordinates \cite{koide20-1}.

\subsection{Stochastic Noether theorem}  \label{sec:s-noether}

From the definition of the probability distribution, the expectation value of the position of the particle is represented by 
\begin{eqnarray}
\int \ud^D x \, x^{i} \rho ({\bf x},t) = \int \ud^D x \, \Psi^* ({\bf x},t)  x^{i} \Psi ({\bf x},t)  \, .
\end{eqnarray}
It is very easy to see that this is consistent with the quantum-mechanical representation.
For the case of the momentum, we have to consider the average of the expectation values of the two velocity fields, 
\begin{eqnarray}
\uM \int \ud^D x \, \frac{ {\bf u}_+ ({\bf x},t) + {\bf u}_- ({\bf x},t) }{2} 
\rho ({\bf x},t) 
&=& 
\int \ud^D x \, \uM {\bf v} ({\bf x},t) \rho ({\bf x},t) \nonumber \\
&=& 
\int \ud^D x \, \Psi^* ({\bf x},t)  (- \ii \hbar \nabla) \Psi ({\bf x},t) \, .
\label{eqn:momentum-exp}
\end{eqnarray}
This coincides with the quantum-mechanical representation.
Consequently, we have to consider the fluctuations for the two velocities fields separately to define the 
standard deviation of momentum as is discussed in Sec.\ \ref{sec:uc-qm}.

It is interesting to notice that 
the above representation of the momentum can be obtained from the stochastic Noether theorem \cite{misawa}.
Let us consider the spatial translation of the stochastic trajectory, 
\begin{eqnarray}
\widehat{\bf r}({\bf R},t) \longrightarrow \widehat{\bf r}^\prime ({\bf R},t) = \widehat{\bf r} ({\bf R},t) + \mbox{\boldmath$\varepsilon$} (\widehat{\bf r} ({\bf R},t),t)\, , 
\end{eqnarray}
where $\mbox{\boldmath$\varepsilon$} ({\bf x},t)$ is an infinitesimal smooth vector function.

As an example, we consider the system described by the following stochastic Lagrangian, 
\begin{eqnarray}
L_{sto} (\widehat{\bf r}, \uD_+ \widehat{\bf r} , \uD_- \widehat{\bf r})
= \frac{\uM}{4} \sum_{l = \pm} (\uD_l \widehat{\bf r}(t))^2 \, .
\end{eqnarray}
Here we set $(\alpha_A,\alpha_B) = (0,1/2)$ and $V({\bf x}) =0$ in the stochastic Lagrangian (\ref{eqn:sto-lag}).
The change of the stochastic action associated with the spatial translation is 
\begin{eqnarray}
I[\widehat{\bf r}^\prime ] - I [\widehat{\bf r} ]
&=& 
\int^{t_f}_{t_i} \ud t \,
\uE \left[
\frac{\partial L_{sto}}{\partial (\uD_+ \widehat{\bf r})}\cdot \uD_+ \mbox{\boldmath$\varepsilon$} 
+ 
\frac{\partial L_{sto}}{\partial (\uD_- \widehat{\bf r})}\cdot \uD_- \mbox{\boldmath$\varepsilon$} 
+ \frac{\partial L_{sto}}{\partial \widehat{\bf r}} \cdot \mbox{\boldmath$\varepsilon$}
\right] \nonumber \\
&=& 
\int^{t_f}_{t_i} \ud t \,
\frac{\ud}{\ud t} \uE \left[
\frac{\partial L_{sto}}{\partial (\uD_+ \widehat{\bf r})}\cdot \mbox{\boldmath$\varepsilon$}
+ 
\frac{\partial L_{sto}}{\partial (\uD_- \widehat{\bf r})}\cdot \mbox{\boldmath$\varepsilon$}
\right]  \nonumber \\
&=& 
\frac{\uM}{2} \int^{t_f}_{t_i} \ud t \,
\frac{\ud}{\ud t} \uE \left[
\uD_+ \widehat{\bf r} ({\bf R},t)\cdot \mbox{\boldmath$\varepsilon$} + \uD_- \widehat{\bf r} ({\bf R},t)\cdot \mbox{\boldmath$\varepsilon$}
\right] \, .
\end{eqnarray}
Here we used that ${\bf u}_\pm ({\bf x},t)$ satisfy the stochastic Euler-Lagrange equation (\ref{eqn:sel}), and 
\begin{eqnarray}
\frac{\ud}{\ud t} \uE[\widehat{X}\widehat{Y}] = \uE[ \widehat{Y}\uD_+ \widehat{X} + \widehat{X} \uD_- \widehat{Y} ] \, ,
\end{eqnarray}
which is obtained from the stochastic partial integration formula (\ref{eqn:spif}).

Suppose that the stochastic action is invariant in the homogeneous translation, 
{\boldmath$\varepsilon$} $\rightarrow const$. 
Then the above equation leads to  
\begin{eqnarray}
\frac{\uM}{2} \frac{\ud}{\ud t} \uE \left[
{\bf u}_+ (\widehat{\bf r} ({\bf R},t),t) 
+ {\bf u}_- (\widehat{\bf r} ({\bf R},t),t)
\right]=0 \, ,
\end{eqnarray}
which is equivalently reexpressed as 
\begin{eqnarray}
\frac{\ud}{\ud t} \int \ud^D x \,  \uM {\bf v} ({\bf x},t) \rho ({\bf x},t)= 0 \, .
\end{eqnarray}
This represents the conservation of the momentum expectation value used in Eq.\ (\ref{eqn:momentum-exp}).

\section{Uncertainty relations for particles in SVM} \label{sec:uc_particle}

Normally the uncertainty relations in quantum mechanics are attributed to the non-commutativity of operators. 
In this section, the same uncertainty relations are shown to be obtained from the non-differentiability of the stochastic trajectory. 
See also the discussion for the uncertainty principle in the original paper by Heisenberg \cite{heisenberg}.
The discussion developed in this section is based on Ref.\ \cite{koide18}.

\subsection{Stochastic Hamiltonian formalism} \label{sec:hami}

In Sec. \ref{sec:s-noether}, 
we found that the first order expectation value of the momentum operator in quantum mechanics 
can be represented by the averaged expectation value of ${\bf u}_\pm ({\bf x},t)$. 
To discuss the uncertainty relations, we should further define the hydrodynamical representation of the fluctuation of momentum. For this purpose, we consider the Hamiltonian formulation of SVM.

The stochastic Hamiltonian is defined through the Legendre transformation of the velocity, 
\begin{eqnarray}
H_{sto} (\widehat{\bf r}, \widehat{\bf p}_+ , \widehat{\bf p}_- )
= 
 \frac{\widehat{\bf p}_+ ({\bf R},t)\cdot \uD_+ \widehat{\bf r}({\bf R},t) + \widehat{\bf p}_- ({\bf R},t)\cdot \uD_- \widehat{\bf r}({\bf R},t)}{2}  
- L (\widehat{\bf r}, \uD_+ \widehat{\bf r},  \uD_- \widehat{\bf r})
\, . \label{eqn:sto_ham}
\end{eqnarray}
Here $\uD_\pm {\bf r}({\bf R},t)$ should be replaced with $\widehat{\bf p}_+({\bf R},t)$ and $\widehat{\bf p}_-({\bf R},t)$ using 
\begin{eqnarray}
\widehat{\bf p}_\pm ({\bf R},t) = 2 \frac{\partial L_{sto}}{\partial (\uD_\pm \widehat{\bf r})} \, . \label{eqn:lad-tra}
\end{eqnarray}
The additional factor $1/2$ in Eq.\ (\ref{eqn:sto_ham}) is introduced for a convention to reproduce the classical result in
the vanishing limit of $\hbar$ (or equivalently $\nu$) \cite{koide18}.

Using the stochastic Lagrangian (\ref{eqn:sto-lag}), the above two momenta are calculated as 
\begin{eqnarray}
\left(
\begin{array}{c}
\widehat{\bf p}_+ ({\bf R},t) \\ 
\widehat{\bf p}_- ({\bf R},t) 
\end{array}
\right)
= 2\uM {\cal M}
\left(
\begin{array}{c}
{\bf u}_+ (\widehat{\bf r}({\bf R},t), t)  \\ 
{\bf u}_- ( \widehat{\bf r}({\bf R},t), t )
\end{array}
\right)
 \, .
\label{eqn:def-momta}
\end{eqnarray}
When the parameter set to reproduce quantum mechanics $(\alpha_A, \alpha_B) = (0,1/2)$ is utilized, 
these are simplified as 
\begin{equation}
\widehat{\bf p}_\pm ({\bf R},t) = \uM {\bf u}_\pm ( \widehat{\bf r}({\bf R},t), t )  \, .
\end{equation}

Substituting these into the stochastic Hamiltonian (\ref{eqn:sto_ham}), we find
\begin{eqnarray}
H_{sto} (\widehat{\bf r}, \widehat{\bf p}_+, \widehat{\bf p}_-)
= 
\frac{1}{8\uM} ( \widehat{\bf p}_+ ({\bf R},t) , \widehat{\bf p}_- ({\bf R},t)  ) 
{\cal M}^{-1} 
\left(
\begin{array}{c}
\widehat{\bf p}_+ ({\bf R},t) \\
\widehat{\bf p}_- ({\bf R},t)
\end{array}
\right) + V(\widehat{\bf r}({\bf R},t)) \, ,
\end{eqnarray}
where
\begin{eqnarray}
{\cal M}^{-1}
=
\frac{4}{4A_+ A_- B^2_+ - B^2_-}
\left(
\begin{array}{cc}
A_- B_+ & -\frac{B_-}{2} \\
-\frac{B_-}{2} & A_+ B_+ 
\end{array}
\right) \, .
\end{eqnarray}
To have the inverse, the parameters $A_\pm$ and $B_\pm$ (or equivalently $\alpha_A$ and $\alpha_B$) 
should satisfy the condition, 
\begin{eqnarray}
{\rm det} ({\cal M} ) \neq 0 \longrightarrow A_+ A_- B^2_+ - \frac{B^2_-}{4} \neq 0\, . \label{eqn:detm0}
\end{eqnarray}

As is analytical mechanics, the variational principle can be formulated with the stochastic Hamiltonian 
and then the stochastic action is defined by  
\begin{eqnarray}
I[\widehat{\bf r}, \widehat{\bf p}_+, \widehat{\bf p}_-] 
= 
\int^{t_f}_{t_i} \ud t \, 
\uE\left[
 \frac{\widehat{\bf p}_+ ({\bf R},t)\cdot \uD_+ \widehat{\bf r} ({\bf R},t)
+ \widehat{\bf p}_- ({\bf R},t)\cdot \uD_- \widehat{\bf r}({\bf R},t)}{2}   
- H_{sto} (\widehat{\bf r}, \widehat{\bf p}_+ , \widehat{\bf p}_- )
\right] \, . \nonumber \\
\end{eqnarray}
We consider the variations of the three quantities, 
\begin{eqnarray}
\begin{split}
\widehat{\bf r} ({\bf R},t) &\longrightarrow \widehat{\bf r}^\prime ({\bf R}.t) = 
\widehat{\bf r} ({\bf R},t) + \delta {\bf f}_{\bf r} (\widehat{\bf r}({\bf R},t), \widehat{\bf p}_+({\bf R},t), \widehat{\bf p}_-({\bf R},t),t) 
\, ,\\
\widehat{\bf p}_\pm ({\bf R},t) &\longrightarrow \widehat{\bf p}^\prime_\pm ({\bf R}.t) = 
\widehat{\bf p}_\pm ({\bf R},t) + \delta {\bf f}_{{\bf p}_\pm} (\widehat{\bf r}({\bf R},t), \widehat{\bf p}_+({\bf R},t), \widehat{\bf p}_-({\bf R},t),t) 
\, .
\end{split}
\end{eqnarray}
where the infinitesimal smooth functions satisfy $\delta{\bf f}_{\bf a} ({\bf x},{\bf y},{\bf z},t_i) =  \delta{\bf f}_{\bf a} ({\bf x},{\bf y},{\bf z},t_f) = 0 \, \, \,  ({\bf a}= {\bf r},{\bf p}_\pm)$.
Then the results of the stochastic variation are summarized by
\begin{equation}
\begin{split}
& \uD_{\pm} \widehat{\bf r}({\bf R},t) = 2 \frac{\partial H_{sto}}{\partial \widehat{\bf p}_\pm } \, , \\
& \left[
\frac{\uD_- \widehat{\bf p}_+ ({\bf R},t)+ \uD_+ \widehat{\bf p}_- ({\bf R},t)}{2} + \frac{\partial H_{sto}}{\partial  \widehat{\bf r}} 
\right]_{\widehat{\bf r}({\bf R},t) ={\bf x}} = 0 \, .
\end{split}
\label{eqn:sto-ham-eq}
\end{equation}
One can easily confirm that the first line of the equations coincides with Eq.\ (\ref{eqn:lad-tra}).
These are the stochastic generalization of the canonical equation.
Note that, to substitute $\widehat{\bf r}({\bf R},t) ={\bf x}$ in the second equation, 
$\widehat{\bf p}_\pm ({\bf R},t)$ are replaced by ${\bf u}_{\pm} (\widehat{\bf r}({\bf R},t),t)$ by using Eq.\ (\ref{eqn:lad-tra}).
Then the stochastic canonical equations are equivalent to the stochastic Euler-Lagrange equation.
As is discussed in Sec.\ \ref{app:bracket}, these results can be expressed by introducing a generalized Poisson bracket.

The parameters $\alpha_A$ and $\alpha_B$ affect only the definitions of the two momenta when the stochastic Hamiltonian does not have a cross term of $\widehat{\bf r}({\bf R},t)$ and $\widehat{\bf p}_\pm({\bf R},t)$. 
In the second equation of the stochastic canonical equation (\ref{eqn:sto-ham-eq}), the two momenta always contribute on an equal footing because of $\partial H_{sto}/\partial  \widehat{\bf r} = \partial V/\partial  \widehat{\bf r}$. 
This property is used to define the standard deviation of momentum later.

A different Hamiltonian formulation is developed in Ref.\ \cite{misawa-hamil}, 
where the symmetrized Lagrangian for $\uD_\pm$ is considered. 
Then only one conjugate momentum associated with $\mathbf{v} (\mathbf{x},t)$ is
introduced while our formulation needs two momenta $\mathbf{p}_\pm({\bf R},t)$.
These two momenta plays a crucial role 
in the construction of the uncertainty relations in hydrodynamics.

The expectation value of the stochastic Hamiltonian gives that of the Hamiltonian operator in quantum mechanics.
From the definition of the stochastic Hamiltonian, we define the following function 
\begin{eqnarray}
&& \mathrm{H}_{sto} ({\bf x},t) 
=
\left.  H_{sto} (\widehat{\bf r},\widehat{\bf p}_+, \widehat{\bf p}_-) \right|_{\widehat{\bf r}({\bf R},t) ={\bf x}}
\nonumber \\
&& \hspace*{0cm}= 
\frac{\uM}{2 {\rm det} ({\cal M})}
\left[
\sum_{l=\pm}
A_l B_+ \left( A_+ A_- B^2_+ + \frac{3B^2_-}{4} \right) {\bf u}^2_l ({\bf x},t)
+ 
B_- \left( 3 A_+ A_- B^2_+ + \frac{B^2_-}{4} \right) {\bf u}_+ ({\bf x},t) \cdot {\bf u}_- ({\bf x},t)
\right] \nonumber \\
&& + V({\bf x}) \, .
\end{eqnarray}
For the parameter $(\alpha_A,\alpha_B, \nu) = (0, 1/2, \hbar/(2\uM))$ which reproduces quantum mechanics, 
this is reduced to
\begin{eqnarray}
 \mathrm{H}_{sto} ({\bf x},t)
= 
\frac{\uM}{4}
\left[
 {\bf u}^2_+ ({\bf x},t) +  {\bf u}^2_- ({\bf x},t)
\right] 
 + V({\bf x}) \, .
\end{eqnarray}
In this case, it is easily confirmed that the expectation value coincides with that of the Hamiltonian operator, 
\begin{eqnarray}
\int \ud^D x \, \rho ({\bf x},t) \mathrm{H}_{sto} ({\bf x},t) 
= \int \ud^D x\, \Psi^* ({\bf x},t) \left[   
-\frac{\hbar^2}{2\uM} \nabla^2 + V({\bf x})
\right] \Psi ({\bf x},t) \, .
\end{eqnarray}
For a general set of parameters, the relation between the stochastic Hamiltonian 
and the energy of the corresponding system is not yet known. 
It is worth mentioning that the canonical equations are not expressed by the standard Poisson bracket in SVM 
and thus the role of Hamiltonian is changed. See the discussion in Sec.\ \ref{app:bracket}.

Note that the quantity corresponding to the Hamiltonian operator is obtained from 
the Bernoulli equation in quantum hydrodynamics. See Appendix \ref{app:ber}.

\subsection{Inequalities in SVM and quantum-mechanical uncertainty relations} \label{sec:uc-qm}

To study uncertainty relations, we have to define the standard deviations of position and momentum. 
For the sake of later convenience, 
we define the following expectation value,
\begin{eqnarray}
\lceil \widehat{f} \, \rfloor = \int \ud^D R \, \rho_0 ({\bf R}) \uE[ f (\widehat{\bf r}({\bf R},t),t) ]
= 
\int \ud^D x \, \rho ({\bf x},t) f ({\bf x},t)
\, , \label{eqn:av_partiacle}
\end{eqnarray} 
where $f({\bf x},t)$ is an arbitrary function and the definition of $\rho ({\bf x},t)$ is given by Eq.\ (\ref{eqn:rho}).

Using this definition, the standard deviation of position is defined by 
\begin{eqnarray}
\left( \sigma^{(2)}_{x^{i}} \right)^{1/2} = \sqrt{ \lceil (\delta \widehat{r}^{i} )^2 \rfloor }\, , \label{eqn:sigma-x-particle}
\end{eqnarray}
where $\delta \widehat{f} =  f (\widehat{\bf r}({\bf R},t),t) - \lceil \widehat{f} \, \rfloor$.
On the other hand, the corresponding definition for momentum is not trivial because now we have two momenta
 $\widehat{\bf p}_\pm ({\bf R},t)$. 
As we noticed, the momenta always contribute on an equal footing in the stochastic canonical equations (\ref{eqn:sto-ham-eq}).
Therefore we define the standard deviation of momentum by 
the average of the two contributions, 
\begin{eqnarray}
\left( \sigma^{(2)}_{p^{i}} \right)^{1/2}=\sqrt{ \frac{\lceil (\delta \widehat{p}^{i}_+ )^2 \rfloor
+ \lceil (\delta \widehat{p}^{i}_-  )^2 \rfloor  }{2} } \, .
\label{delta_p}
\end{eqnarray}
To calculate this expectation value, we should remember that the above momenta are functions of $\widehat{\bf r}({\bf R},t)$.
It is easily shown that this definition coincides with the quantum-mechanical one \cite{koide18}.
See Appendix \ref{app:relationtoqm}.

The product of the above two deviations is reexpressed as
\begin{eqnarray}
\sigma^{(2)}_{x^{i}} \sigma^{(2)}_{p^{j}} 
= 
\sigma^{(2)}_{x^{i}} \left\{
\left\lceil  ( \delta \widehat{p}^j_\um)^2 \right\rfloor + \left\lceil (\delta \widehat{p}^j_{\ud} )^2 \right\rfloor
 \right\}
\equiv
\sigma^{(2)}_{x^{i}} \sigma^{(2)}_{p^j_\um}  + \sigma^{(2)}_{x^{i}} \sigma^{(2)}_{p^j_\ud} 
  \, , \label{eqn:ss}
\end{eqnarray}
where
\begin{equation}
\begin{split}
\widehat{\bf p}_\um ({\bf R},t) &= \frac{\widehat{\bf p}_+({\bf R},t) + \widehat{\bf p}_-({\bf R},t) }{2} \, ,\\
\widehat{\bf p}_{\ud} ({\bf R},t) &=  \frac{\widehat{\bf p}_+({\bf R},t) - \widehat{\bf p}_-({\bf R},t) }{2} \, .
\end{split}
\end{equation}
From the consistency condition given by Eq.\ (\ref{eqn:cc}), these are reexpressed as
\begin{equation}
\begin{split}
\widehat{\bf p}_\um ({\bf R},t)
&= 
\uM \left\{ {\bf v} ( \widehat{\bf r}({\bf R},t),t) ) + 2 \nu \alpha_A B_+ \nabla \ln \rho (\widehat{\bf r}({\bf R},t),t) 
\right\}
\, , \\
\widehat{\bf p}_\ud ({\bf R},t)  
&= 
\uM \left\{
 2\nu \alpha_B \nabla \ln \rho (\widehat{\bf r}({\bf R},t),t) + 2\alpha_A B_+ {\bf v} (\widehat{\bf r}({\bf R},t),t) 
\right\} \, .
\end{split}
\end{equation}
The terms proportional to $\nabla \ln \rho  (\widehat{\bf r}({\bf R},t),t)$ contribute to 
a finite minimum value of the uncertainty relations. 

Applying the Cauchy-Schwarz inequality $\uE[\widehat{A}^2]\uE[\widehat{B}^2] \ge |\uE[\widehat{A}\widehat{B}]|^2$, the first and second terms on the right-hand side of Eq.\ (\ref{eqn:ss}) satisfy 
\begin{equation}
\begin{split}
\sigma^{(2)}_{x^{i}} \sigma^{(2)}_{p^j_\um} 
&\ge 
\left|
\left\lceil \delta \widehat{r}^{i} \,  \delta \widehat{p}^j_{\um} \, \right\rfloor
\right|^2 
= 
\uM^2 \left|
- 2\nu \alpha_A B_+ \delta_{ij} + \lceil \delta \widehat{r}^{i} \, \delta \widehat{v}^{j}  \rfloor
\right|^2
\, , \\
\sigma^{(2)}_{x^{i}} \sigma^{(2)}_{p^j_\ud} 
&\ge 
\left|
\left\lceil \delta \widehat{r}^{i} \,  \delta \widehat{p}^j_{\ud}\, \right\rfloor
\right|^2
= 
\uM^2 \left| - 2\nu \alpha_B \delta_{ij}  +2 \alpha_A B_+  \lceil \delta \widehat{r}^{i}  \, \delta \widehat{v}^{j} \rfloor \right|^2 \, ,
\end{split}
\end{equation}
respectively.
In this derivation we used 
\begin{equation}
\begin{split}
\lceil \widehat{r}^{i} \widehat{p}^{j}_{\um}\, \rfloor
&=
- 2\uM \nu \alpha_A B_+ \delta_{ij} + \uM  \int \ud^D x \, \rho ({\bf x},t) x^{i} v^{j} ({\bf x},t) \, ,\\
\lceil \widehat{r}^{i} \widehat{p}^{j}_{\ud}\, \rfloor
&=
- 2\uM \nu \alpha_B \delta_{ij} + 2\uM \alpha_A  B_+  \int \ud^D x \, \rho ({\bf x},t) x^{i} v^{j} ({\bf x},t) \, .
\end{split}
\label{eqn:xpdm-particle}
\end{equation}

Substituting these results into Eq.\ (\ref{eqn:ss}), 
the inequality satisfied for position and momentum is given by \cite{koide18}
\begin{eqnarray}
\sigma^{(2)}_{x^{i}} \sigma^{(2)}_{p^{j}}  
\ge 
\uM^2 \left[ \frac{(4\nu \alpha_A^2 B^2_+ - \nu (B_+ - B_-))^2}{1 + 4 \alpha^2_A B^2_+} \delta_{ij} 
+ (1 + 4 \alpha^2_A B^2_+) 
\left(
\lceil \delta \widehat{r}^{i}  \, \delta \widehat{v}^{j}  \rfloor
- \frac{4\nu \alpha_A B^2_+ }{1 + 4 \alpha^2_A B^2_+}\delta_{ij}
\right)^2 \right] 
\, . 
\label{eqn:inequality-particle1}
\nonumber \\
\end{eqnarray}
This is the most general representation of the uncertainty relation for single-particle systems in SVM.

This reproduces the standard uncertainty relations in quantum mechanics.
Choosing $(\alpha_A,\alpha_B,\nu) = ( 0, 1/2, \hbar/(2\uM))$, 
the above inequality is reduced to
\begin{eqnarray}
\left( \sigma^{(2)}_{x^{i}} \right)^{1/2} \left( \sigma^{(2)}_{p^{j}} \right)^{1/2} 
&\ge& 
 \sqrt{
\left( \frac{\hbar}{2} \right)^2 \delta_{ij} 
+
\uM^2 \left(
\lceil \delta \widehat{r}^{i}  \, \delta \widehat{v}^{j}  \rfloor
\right)^2 } \nonumber \\
&=&
\sqrt{ \left( \frac{\hbar}{2} \right)^2 \delta_{ij}
+ ( {\rm Re} [
\langle (x_{op}^{i} - \langle x_{op}^{i} \rangle) (p_{op}^{j} - \langle p_{op}^{j} \rangle) \rangle 
] )^2
}
\, .\label{eqn:rs-ineq}
\end{eqnarray}
In the second line, $\langle~~~~\rangle$ means the quantum mechanical expectation value with a wave function.
The position and the momentum operators in Cartesian coordinates are denoted 
by $x_{op}^{i}$ and $p_{op}^{j}$, respectively.
This inequality is known as the Robertson-Schr\"{o}dinger inequality in quantum mechanics. 
When the second term inside the square root on the right-hand side is ignored, this inequality coincides with the Kennard inequality,
\begin{eqnarray}
\left( \sigma^{(2)}_{x^{i}} \right)^{1/2} \left( \sigma^{(2)}_{p^{j}} \right)^{1/2} 
\ge 
\frac{\hbar}{2}  \delta_{ij}
\, .
\end{eqnarray}
In addition, one can easily confirm in the Robertson-Schr\"{o}dinger inequality that the second term inside the square root vanishes for the coherent state.

For the case of the Kennard inequality, the right-hand side of Eq.\ (\ref{eqn:inequality-particle1}) is simplified as 
\begin{eqnarray}
\left( \sigma^{(2)}_{x^{i}} \right)^{1/2} \left( \sigma^{(2)}_{p^{j}} \right)^{1/2} 
\ge 
\uM \nu \frac{|4 \alpha_A^2 B^2_+ - (B_+ - B_-)|}{\sqrt{1 + 4 \alpha^2_A B^2_+}} \delta_{ij} 
= \frac{4\uM\nu}{\sqrt{1 + 4 \alpha_A^2 B^2_+}} |{\rm det} ({\cal M})| \delta_{ij} \, .
\end{eqnarray}
From this representation, one can easily see that 
the minimum value of uncertainty never vanish for a finite $\nu$ 
because ${\rm det} ({\cal M}) \neq 0$. 
That is, a finite minimum uncertainty for position and momentum 
is not inherent in quantum mechanics but universal in stochastic dynamics formulated in SVM.

The Cauchy-Schwarz inequality is a special case of the H\"{o}lder inequality, 
\begin{eqnarray}
\left( \uE[|\widehat{A}|^a] \right)^{1/a} \left( \uE[|\widehat{B}|^b] \right)^{1/b} \ge \uE[|\widehat{A}\widehat{B}|] \ge |\uE[\widehat{A}\widehat{B}]| \, ,
\end{eqnarray}
where the positive real $a$ and $b $ satisfy $1< a,b$ and $1/a+1/b = 1$. The Cauchy-Schwarz inequality is reproduced for $a=b=2$.
Using this, we can study a different inequality satisfied for quantum mechanics.
Let us introduce 
\begin{equation}
\begin{split}
\sigma^{(a)}_{x^{i}} &= \lceil |\delta \widehat{r}^{i}|^a \rfloor  \, , \\
\sigma^{(b)}_{p^{i}} &= \frac{\lceil |\delta \widehat{p}^{i}_+|^b \rfloor + \lceil |\delta \widehat{p}^{i}_-|^b \rfloor }{2} \, .
\end{split}
\end{equation}
The product of this quantities satisfies the same uncertainty relation as before, 
\begin{eqnarray}
\lefteqn{\left( \sigma^{(a)}_{x^{i}} \right)^{1/a} \left( \sigma^{(b)}_{p^{i}} \right)^{1/b} } && \nonumber \\
&\ge& 
\uM \sqrt{\frac{(4\nu \alpha_A^2 B^2_+ - \nu (B_+ - B_-))^2}{1 + 4 \alpha^2_A B^2_+} \delta_{ij} 
+ (1 + 4 \alpha^2_A B^2_+) 
\left(
\lceil \delta \widehat{r}^{i}  \, \delta \widehat{v}^{j}  \rfloor
- \frac{4\nu \alpha_A B^2_+ }{1 + 4 \alpha^2_A B^2_+}\delta_{ij}
\right)^2 
  }\, . 
\nonumber \\
\end{eqnarray}

The consistency condition (\ref{eqn:cc}) is the origin of the finite minimum uncertainty in SVM 
as was shown in the above derivation.
In fact, the consistency condition has a close relation to the canonical commutation rule in the standard formulation of quantum mechanics. 
Using the definitions of the two momenta (\ref{eqn:def-momta}), the consistency condition is reexpressed as 
\begin{eqnarray}
\left\lceil \frac{1}{2} (\widehat{r}^i \widehat{p}^j_- - \widehat{r}^i \widehat{p}^j_+ ) 
+ \alpha_A B_+ (\widehat{r}^i \widehat{p}^j_- + \widehat{r}^i \widehat{p}^j_+ ) 
\right\rfloor
= 4\uM \nu {\rm det (\cal M)} \delta^{ij}\, .
\label{eqn:cc_pgeral}
\end{eqnarray} 
For the parameter set to reproduce the Schr\"{o}dinger equation, $(\alpha_A,\alpha_B,\nu) = (0,1/2,\hbar/(2\uM) )$, this is simplified as 
\begin{eqnarray}
\left\lceil  \, \widehat{r}^i \widehat{p}^j_- - \widehat{r}^i \widehat{p}^j_+ \, 
\right\rfloor
= \hbar \delta^{ij} \, . \label{eqn:cc_ppm}
\end{eqnarray}
This reminds us the canonical commutation rule although the imaginary unit ($\ii$) does not appear in the above equation. 
The role of the imaginary unit and a more detailed discussion will be developed in a forthcoming paper.

\subsection{Generalized bracket} \label{app:bracket}

The canonical equations in analytical mechanics can be expressed with the Poisson bracket. 
The analogous representation seems to be possible in SVM by introducing the generalized bracket,
\begin{eqnarray}
\{ X_1 , X_2 | Y_+ , Y_- \} 
&=& 
\sum_{i=1}^D \left(
\frac{\partial X_1}{\partial \widehat{r}^{i}} \frac{\partial Y_+}{\partial \widehat{p}^{i}_+ }
+
\frac{\partial X_2}{\partial \widehat{r}^{i}} \frac{\partial Y_-}{\partial \widehat{p}^{i}_- }
- 
\frac{\partial X_1}{\partial \widehat{p}^{i}_- }\frac{\partial Y_-}{\partial \widehat{r}^{i}} 
-
\frac{\partial X_2}{\partial \widehat{p}^{i}_+ }\frac{\partial Y_+}{\partial \widehat{r}^{i}} 
\right) \, . 
\end{eqnarray}
This bracket is invariant for the simultaneous exchange between $(X_1, Y_+, \widehat{p}_+)$ and 
$(X_2, Y_-, \widehat{p}_-)$.

We further suppose that the matrix ${\cal M}$ does not have off-diagonal components (that is, $\alpha_B = 1/2$) and hence the stochastic Hamiltonian does not have a cross term of  
$\widehat{\bf p}_+ ({\bf R},t)$ and $\widehat{\bf p}_- ({\bf R},t)$.
Then the stochastic Hamiltonian can be decomposed into the two parts as 
\begin{eqnarray}
H_{sto} (\widehat{\bf r}, \widehat{\bf p}_+ , \widehat{\bf p}_- ) = H^{(+)} (\widehat{\bf r}, \widehat{\bf p}_+ )+ H^{(-)} (\widehat{\bf r},  \widehat{\bf p}_- ) \, .
\label{eqn:decom-ham}
\end{eqnarray}
This decomposition has an ambiguity for adding a constant but this is irrelevant in the following discussion. 
Then the stochastic canonical equations (\ref{eqn:sto-ham-eq}) are formally expressed using the above bracket as
\begin{equation}
\begin{split}
&  \frac{\uD_+ \widehat{r}^{i}  + \uD_- 0}{2} =\frac{1}{2} \uD_+ \widehat{r}^{i}  =  \{ \widehat{r}^{i}, 0 |  H^{(+)},  H^{(-)} \} \, ,\\
&  \frac{\uD_+ 0 + \uD_- \widehat{r}^{i} }{2} = \frac{1}{2} \uD_- \widehat{r}^{i}  =  \{ 0, \widehat{r}^{i}  |  H^{(+)},  H^{(-)} \} \, ,\\
 & \left[ \frac{\uD_+ \widehat{p}^{i}_- + \uD_- \widehat{p}^{i}_+ }{2}  - \{ \widehat{p}^{i}_- , \widehat{p}^{i}_+ |  H^{(+)},  H^{(-)} \} \right]_{\widehat{\bf r}({\bf R},t)= {\bf x}} = 0\, .
\end{split}
\end{equation}
It is however noted that the applicability of the generalized bracket is not 
confirmed for a general function, like 
$f (\widehat{\bf r}^{i}({\bf R},t), \widehat{\bf p}^{i}_+ ({\bf R},t), \widehat{\bf p}^{i}_- ({\bf R},t))$.

Another generalization of the Poisson bracket is known in classical mechanics by Nambu where the generalized bracket depends on three variables \cite{nambu}. 
His approach however will be difficult to apply to our case because of the two different time derivatives in SVM.

\section{Uncertainty Relations for continuum media} \label{sec:uc_fluid}

We have discussed the quantization of single-particle systems in SVM, showing that quantization corresponds to the stochastic variation of a classical action. 
The applicability is however not limited to quantization. 
Viscous hydrodynamics can be formulated in SVM.

\subsection{Brief summary of variational approach to ideal fluid} \label{sec:idealfluid_vari}

\begin{figure}[t]
\begin{center}
\includegraphics[scale=0.4]{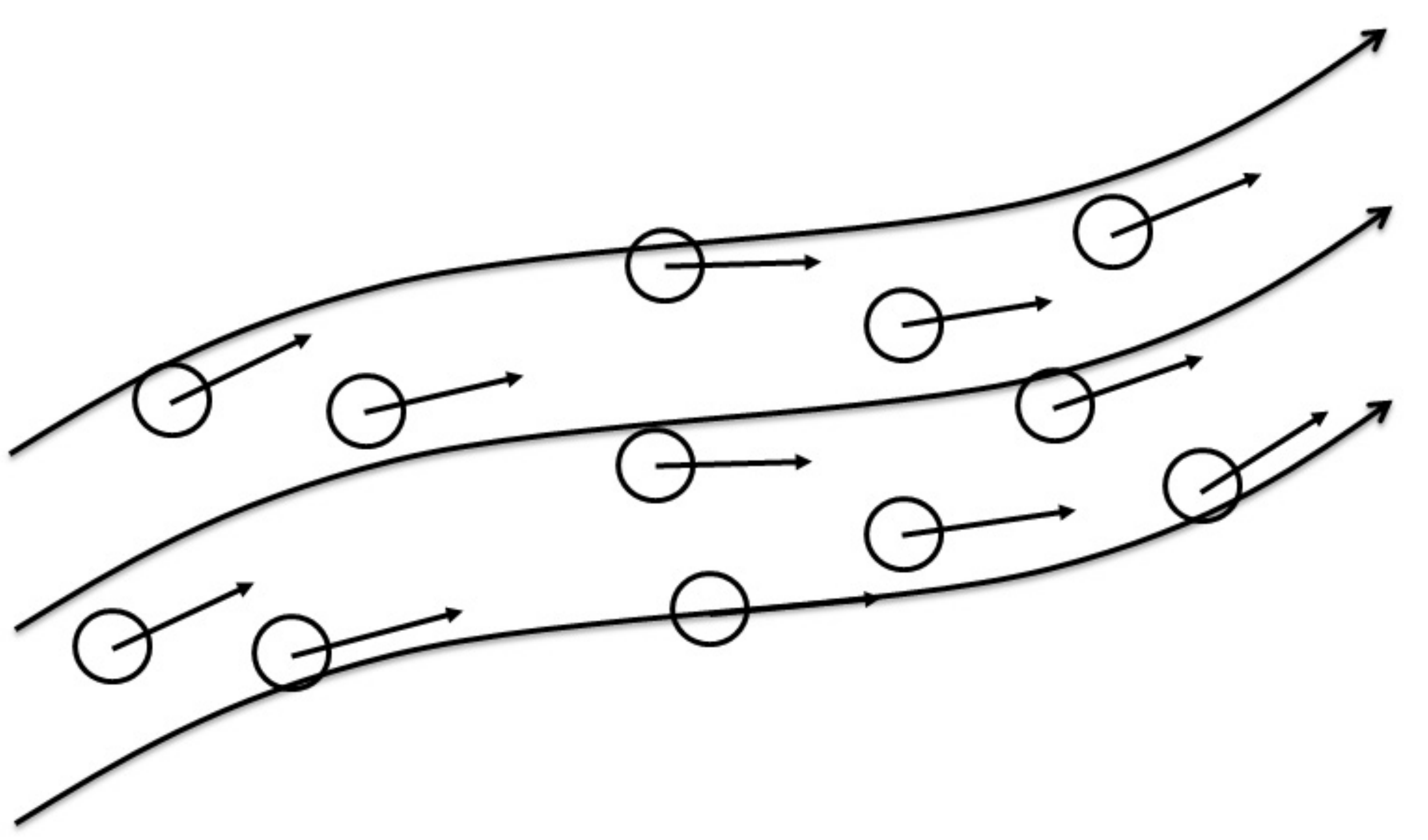}
\end{center}
\caption{The behavior of fluid can be understood as the ensemble of fluid elements.}
\label{fig:euler-lagrange}
\end{figure}

We consider the ideal fluid as a continuum medium which is described by the mass distribution $n({\bf x},t)$ 
and the velocity field ${\bf v} ({\bf x},t)$.
The ideal-fluid action is 
\begin{eqnarray}
I [\rho, {\bf v}, \lambda] = \int^{t_f}_{t_i} \ud t \int \ud^D x \, {\cal L} ({\bf x},t)\, , \label{eqn:idealaction1}
\end{eqnarray}
where the Lagrangian density is defined by 
\begin{eqnarray} 
{\cal L} ({\bf x},t) = \frac{n({\bf x},t)}{2} {\bf v}^2({\bf x},t) - \varepsilon (n({\bf x},t)) 
+\lambda({\bf x},t) [\partial_t n({\bf x},t) + \nabla \cdot \{n({\bf x},t) {\bf v}({\bf x},t) \} ]
\, . \label{eqn:ideal_eular_lag}
\end{eqnarray}
The internal energy density $\varepsilon$ is assumed to be a function only of $n({\bf x},t)$.
The mass distribution is normalized by 
\begin{eqnarray}
\uM_T = \int \ud^D x \, n({\bf x},t) \, ,
\end{eqnarray}
where $\uM_T$ is the total mass of the fluid.
The first and the second terms on the right-hand side of Eq.\ (\ref{eqn:ideal_eular_lag}) correspond to the kinetic and potential terms, respectively. 
The third term represents the constraint of the mass conservation with the Lagrangian multiplier $\lambda ({\bf x},t)$.

There is, however, another view to describe fluids. 
Fluids can be represented by the ensemble of fluid elements which are virtual particles with fixed masses,  
as is shown in Fig.\ \ref{fig:euler-lagrange}.
The size of a fluid element is infinitesimal compared to the macroscopic scale of observation, but 
sufficiently large so that constituent particles in each element are thermally equilibrated.
This is known as the assumption of 
the local thermal equilibrium and a fundamental requirement in hydrodynamics.

The trajectory of the fluid element is denoted by ${\bf r}({\bf R},t)$ where ${\bf R}$ is 
the initial position of fluid elements, ${\bf r}({\bf R},t_i) = {\bf R}$.
Then the mass distribution is expressed as 
\begin{eqnarray}
n ({\bf x},t) = \int \ud^D R \,  n_{0} ({\bf R}) \delta^{(D)} ({\bf x}- {\bf r}({\bf R},t)) \, , \label{eqn:n_lag}
\end{eqnarray}
where $n_{0} ({\bf R})$ is the initial mass distribution satisfying 
\begin{eqnarray}
\uM_T = \int \ud^{D} R \, n_{0} ({\bf R}) \, .
\end{eqnarray}
From the above definition of $n({\bf x},t)$, the equation of continuity for the fluid mass is automatically satisfied,
\begin{eqnarray}
\partial_t n({\bf x},t) 
= - \nabla \cdot \{ n({\bf x},t) {\bf v}({\bf x},t) \} \, ,
\end{eqnarray}
where we used the identification 
\begin{eqnarray}
\frac{\ud {\bf r}({\bf R},t)}{\ud t} = {\bf v} ({\bf r}({\bf R},t) ,t) \, .\label{eqn:lag-identi}
\end{eqnarray}

The action for the ideal fluid in terms of fluid elements is reexpressed as 
\begin{eqnarray}
I [{\bf r}] 
= \int^{t_f}_{t_i} \ud t \int \ud^D R \, n_0 ({\bf R}) 
\left[
\frac{1}{2} \left( \frac{\ud {\bf r}({\bf R},t)}{\ud t} \right)^2 
- J ({\bf r}) \frac{\varepsilon (n_0 ({\bf R})/J ({\bf r}))}{n_0 ({\bf R})}
\right] \, . \label{eqn:idealaction2}
\end{eqnarray}
To obtain this, note that 
\begin{eqnarray}
J ({\bf r}) n ({\bf r}({\bf R},t),t) = n_0 ({\bf R}) \, ,
\end{eqnarray}
where the Jacobian is defined by 
\begin{eqnarray}
J ({\bf r}) = {\rm det} \left( \frac{\partial {\bf r}({\bf R},t)}{ \partial {\bf R}} \right) \, . 
\end{eqnarray}
We find that the variable of variation in this action (\ref{eqn:idealaction2}) is given only by ${\bf r}({\bf R},t)$, while 
$(\rho,{\bf v},\lambda)$ are chosen as the variables in Eq.\ (\ref{eqn:idealaction1}).
This is because the mass distribution and the velocity field are expressed using ${\bf r}({\bf R},t)$ and its time derivative, 
and the mass conservation is automatically satisfied in Eq.\ (\ref{eqn:n_lag}).

We can implement the variation as is the case in single-particle systems.
The variation of the trajectory of fluid elements is given by
\begin{eqnarray}
{\bf r}({\bf R},t) \longrightarrow {\bf r}^\prime ({\bf R},t) = {\bf r}({\bf R},t) + \delta {\bf f} ({\bf r}({\bf R},t),t) \, ,
\end{eqnarray}
where the infinitesimal smooth function $\delta {\bf f} ({\bf x},t)$ satisfies the same boundary condition introduced before, 
$\delta {\bf f} ({\bf x},t_i) = \delta {\bf f} ({\bf x},t_f) = 0$.
The variation of the potential term of the action is calculated as
\begin{eqnarray}
\lefteqn{\delta  \int^{t_f}_{t_i} \ud t \int \ud^D R \, n_0 ({\bf R}) J ({\bf r}) \frac{\varepsilon (n_0 ({\bf R})/J ({\bf r}))}{n_0 ({\bf R})}} && \nonumber \\
&=& 
\int^{t_f}_{t_i} \ud t \int \ud^D R \, n_0 ({\bf R}) \left( \frac{\ud}{\ud n} 
\frac{\varepsilon (n) }{n} \right) \delta \frac{n_0 ({\bf R})}{J({\bf r})} \nonumber \\
&=& 
 \int^{t_f}_{t_i} \ud t \int \ud^D R \, \left[\frac{n_0 ({\bf R})}{n({\bf r}({\bf R},t),t)} 
\nabla P (n({\bf r}({\bf R},t),t)) \right] \cdot  \delta {\bf f} ({\bf r}({\bf R},t),t) \, ,
\end{eqnarray}
where the adiabatic (isentropic) pressure is thermodynamically defined as 
\begin{eqnarray}
\ud E = - P \ud V \longrightarrow P = - \frac{\ud}{ \ud n^{-1}} \left( \frac{\varepsilon}{n} \right)\, .
\end{eqnarray}
From the second to the third line, we used 
\begin{eqnarray}
\delta \frac{n_0 ({\bf R})}{J({\bf r})}
= 
- \frac{n({\bf r}({\bf R},t),t)}{J({\bf r})} \sum_{i,j=1}^D A_{ij} \frac{\partial \delta f^i ({\bf r}({\bf R},t),t)}{\partial R^j} 
\, .
\end{eqnarray}
The cofactor of the Jacobian is defined by   
\begin{eqnarray}
A_{ij} = \frac{\partial J ({\bf r})}{\partial (\partial r^i/\partial R^j)} \, ,
\end{eqnarray}
which satisfies 
\begin{equation}
\begin{split}
\sum_{j=1}^D \frac{\partial A_{ij}}{\partial R^j} &= 0 \, , \\
\sum_{j=1}^D A_{ij} \frac{\partial}{\partial R^j} &= J ({\bf r}) \partial_i \, .
\end{split}
\end{equation}

The variation of the action eventually leads to 
\begin{eqnarray}
&& \delta I [{\bf r}] \nonumber \\
&=& 
\int^{t_f}_{t_i} \ud t \int \ud^D R \, n_0 ({\bf R}) 
\left[
\frac{\ud}{\ud t} {\bf v} ({\bf r}({\bf R},t),t) + \frac{1}{n({\bf r}({\bf R},t),t)} 
\nabla P (n({\bf r}({\bf R},t),t))
\right] \cdot \delta {\bf f} ({\bf r}({\bf R},t),t) \nonumber \\
&=& 
\int^{t_f}_{t_i} \ud t \int \ud^D x \, n ({\bf x},t)
\left[
 (\partial_t + {\bf v}({\bf x},t)\cdot \nabla ){\bf v}({\bf x},t) + \frac{1}{n({\bf x},t)} \nabla P (n({\bf x},t))
\right] \cdot \delta {\bf f} ({\bf r}({\bf R},t),t) \, . \nonumber \\
\end{eqnarray}
From the Hamilton principle, the velocity field should be given by the solution of the 
following equation, 
\begin{eqnarray}
(\partial_t + {\bf v}({\bf x},t)\cdot \nabla ){\bf v}({\bf x},t) = - \frac{1}{n({\bf x},t)} \nabla P (n({\bf x},t)) \, .\label{eqn:euler_eq}
\end{eqnarray}
This is the Euler equation for the ideal fluid.
The same argument is applicable to derive the relativistic Euler equation interacting with electromagnetic fields \cite{hugo}.

To show the equivalence between the two actions (\ref{eqn:idealaction1}) and (\ref{eqn:idealaction2}), 
we need to assume that the trajectories of different fluid elements are not across. 
Such a condition will, however, not be satisfied in the case of turbulence.

\subsection{Derivation of compressible NSF equation in SVM} \label{sec:cnsf}

In the derivation of the Euler equation, 
we consider the variation only of the smooth trajectory of the fluid element. 
Now we generalize it and take into account the variation of the non-differentiable trajectory following the 
discussion developed in the previous section \cite{koide12,koide18,koide20-2}.
There are many works to derive the NSF equation of the incompressible fluid 
in the different formulations of the stochastic calculus of variations \cite{inoue,nakagomi,marra,gomes,eyink,del,nov,yasue-ns,cruzeiro}.
The applications to the compressible fluid are developed exclusively using the framework presented in this paper \cite{koide12,koide18,koide20-2}.

The application of SVM to a continuum medium is straightforward.
The zigzag trajectory of fluid elements is characterized by the forward and backward SDE's which are defined by  
\begin{equation}
\begin{split}
\ud \widehat{\bf r}({\bf R},t) &=  {\bf u}_+ (\widehat{\bf r}({\bf R},t), t) \ud t + \sqrt{2\nu} \ud\widehat{\bf W}(t)  \, \, \, (\ud t > 0) \, 
, \\
\ud \widehat{\bf r}({\bf R},t) &=  {\bf u}_- (\widehat{\bf r}({\bf R},t), t) \ud t + \sqrt{2\nu} \ud \underline{\widehat{\bf W}}(t)  \, \, \, (\ud t < 0) \, ,
\end{split}
\end{equation} 
respectively. The standard Wiener processes are the same as those in the single-particle systems.

The definition of the mass distribution is modified by operating the ensemble average for the Wiener processes, 
\begin{eqnarray}
n ({\bf x},t) = \int \ud^D R \,  n_{0} ({\bf R}) \uE[ \delta^{(D)} ({\bf x}- {\bf r}({\bf R},t)) ] \, ,\label{eqn:}
\end{eqnarray}
and then the corresponding consistency condition is expressed with the mass distribution as 
\begin{eqnarray}
 {\bf u}_+ ({\bf x},t) =  {\bf u}_- ({\bf x},t) + 2\nu \nabla \ln n ({\bf x},t) 
\label{eqn:cc2}
\, .
\end{eqnarray}
Using the consistency condition, the equation for the mass distribution is shown to be given by the equation of continuity, 
\begin{eqnarray}
\partial_t n ({\bf x},t) = - \nabla \cdot \{ n({\bf x},t) {\bf v}({\bf x},t) \} \, ,
\end{eqnarray}
where the fluid velocity is given by the average of ${\bf u}_\pm ({\bf x},t)$ as is defined in Eq.\ (\ref{eqn:ave-vel}).

The stochastic action corresponding to Eq.\ (\ref{eqn:idealaction2}) is defined by 
\begin{eqnarray}
I_{sto} [\widehat{\bf r}] 
= \int^{t_f}_{t_i} \ud t \int \ud^D R \, \frac{n_0 ({\bf R})}{\uM} \uE[L_{sto}(\widehat{\bf r},\uD_+ \widehat{\bf r}, \uD_- \widehat{\bf r})]\, , \label{eqn:sto_ac_fluid}
\end{eqnarray}
where the stochastic Lagrangian is given by
\begin{eqnarray}
\lefteqn{L_{sto}(\widehat{\bf r},\uD_+ \widehat{\bf r}, \uD_- \widehat{\bf r})
} && \nonumber \\
&=& 
\uM
\left[
\frac{1}{2} (\uD_+ \widehat{\bf r}({\bf R},t), \uD_- \widehat{\bf r}({\bf R},t)) 
{\cal M} 
\left(
\begin{array}{c}
\uD_+ \widehat{\bf r}({\bf R},t) \\
\uD_- \widehat{\bf r}({\bf R},t)
\end{array}
\right) 
- J(\widehat{\bf r}) \frac{\varepsilon ( n_0 ({\bf R})/J(\widehat{\bf r})  ) }{n_0 ({\bf R})}
\right]
\, .
\label{eqn:lag-con-me}
\end{eqnarray}
The matrix ${\cal M}$ is defined in Eq.\ (\ref{eqn:cal_m}).
For the consistency of the dimension of the Lagrangian, we introduced a mass parameter $\uM$.
It should be emphasized that the internal energy density $\varepsilon$ can be a function not only of the mass distribution but also, 
for example, of the entropy density. 
This is different from the case of the ideal fluid where the entropy does not change along the motion of fluid element. In the following calculation, 
however, we assume that the contribution from the entropy density in variation are negligibly small and 
thus the entropy dependence in the internal energy density is omitted. This simplification, however, affects the definition of the second coefficient of viscosity. See the discussion below Eq.\ (\ref{eqn:coef_nsf}).

Applying the procedure developed in the previous section to Eq.\ (\ref{eqn:lag-con-me}),
 the stochastic Hamilton principle leads to 
\begin{eqnarray}
\left(\partial_t + {\bf u}_- ({\bf x},t) \cdot \nabla - \nu \nabla^2, \,  \partial_t + {\bf u}_+ ({\bf x},t) \cdot \nabla + \nu \nabla^2 \right)
{\cal M} 
\left(
\begin{array}{c}
{\bf u}_+ ({\bf x},t ) \\
{\bf u}_- ({\bf x},t ) 
\end{array}
\right)
= - \frac{1}{n({\bf x},t)}\nabla P(n ({\bf x},t)) \, . \label{eqn:vari-con-med1}
\nonumber \\
\end{eqnarray}
Representing the left-hand side using the fluid velocity ${\bf v}({\bf R},t)$, we find 
\begin{eqnarray}
n ({\bf x},t) (\partial_t + {\bf v}({\bf x},t)\cdot \nabla ) v^{i} ({\bf x},t) 
-  
\sum_{j=1}^D \partial_j \{ \eta  E^{ij} ({\bf x},t)+ \kappa \Pi^{ij}_Q({\bf x},t) \}
= -  \partial_i \{ P(n({\bf x},t)) - \zeta \theta ({\bf x},t) \} \, , \nonumber \\\label{nsf.eq}
\end{eqnarray}
where 
\begin{equation}
\begin{split}
E^{ij} ({\bf x},t) &= \frac{1}{2} ( \partial_i v^j ({\bf x},t) + \partial_j v^i ({\bf x},t)) - \frac{1}{D} \theta ({\bf x},t) \delta_{ij} \, , \\
\theta ({\bf x},t) &= \nabla \cdot {\bf v} ({\bf x},t) \, ,\\
\Pi^{ij}_Q ({\bf x},t) &=  n({\bf x},t) \partial_i \partial_j \ln n({\bf x},t)  \, ,
\end{split}
\end{equation}
and the coefficients are defined by 
\begin{equation}
\label{eqn:coef_nsf}
\begin{split}
\kappa &= 2 \alpha_B \nu^2\, ,\\
\eta &= 2 \alpha_A (1 + 2 \alpha_B) \nu n ({\bf x},t) \, ,\\
\zeta &= \frac{\eta}{D} \, .
\end{split}
\end{equation}
The shear viscosity term and the bulk viscosity term are represented by $\eta E^{ij} ({\bf x},t)$ and 
$\zeta \theta({\bf x},t)$, respectively.
This equation is reduced to the Navier-Stokes-Fourier (NSF) equation when the $\Pi^{ij}_Q $ term  
is dropped by setting $\kappa  = 0$ ($\alpha_B = 0$).
Note however that the finite contribution of the so-called second coefficient of viscosity is not reproduced 
in the above derivation and thus the bulk viscosity $\zeta$ is simply proportional to the shear viscosity $\eta$ as is shown in the last equation of Eq.\ (\ref{eqn:coef_nsf}). 
To reproduce this second coefficient, we should consider, for example, the variation of the entropy dependence 
in the pressure, as is discussed in Ref.\ \cite{koide12}.
In the following discussion, however, 
we ignore this effect.

The last term on the left-hand side, $\Pi^{ij}_Q ({\bf x},t)$, corresponds to the quantum potential, because
\begin{eqnarray}
\partial_i \frac{\nabla^2 \sqrt{n({\bf x},t)}}{\sqrt{n({\bf x},t)}} 
= \frac{1}{2n({\bf x},t)} \nabla \cdot \left\{ n({\bf x},t) \nabla\partial_i  \ln n({\bf x},t) \right\} \,.
\end{eqnarray}
Differently from quantum hydrodynamics (\ref{eqn:vari_qh}), however, the above term is induced by thermal fluctuations, not by quantum fluctuations.

The emergence of the $\Pi^{ij}_Q ({\bf x},t)$ term reminds us the possible modification of the NSF equation proposed by Brenner \cite{brenner1,brenner2,klimontovich,ottinger,graur,greensh,eu,don,dadzie,gustavo,laksh2019}. 
Normally, the fluid velocity is defined so as to be parallel to the mass flow.
Brenner pointed out that the velocity of a tracer particle of fluids is not necessarily parallel to such a velocity and 
the existence of these two velocities should be taken into account in the formulation of hydrodynamics. 
Then, by applying the linear irreversible thermodynamics, 
the difference of the two velocities is found to be characterized 
by the gradient of $n({\bf x},t)$ as is given by our consistency condition (\ref{eqn:cc}) and a new effect corresponding to the $\Pi^{ij}_Q$ term appears in the modified hydrodynamics. 
See also the Table 1 in Ref.\ \cite{brenner2}.
This theory is called bivelocity hydrodynamics but the introduction of such a modification is still controversial. 
The similar situation is well-known in the community of relativistic hydrodynamics where two different fluid velocities are introduced through the energy flow (Landau-Lifshitz velocity) and a conserved charge flow (Eckart velocity) \cite{kodama-review}.
In fact, the structure analogous to bivelocity hydrodynamics 
naturally appears as the next-to-leading order relativistic corrections to the NSF equation \cite{gustavo}.
Let us denote the Landau-Lifshitz velocity   
and the Eckart velocity by ${\bf u}_L ({\bf x},t)$ and ${\bf u}_E ({\bf x},t)$, respectively. 
Using Eq.\ (37) of Ref.\ \cite{gustavo} and Fick's law of diffusion, we find that these two velocities satisfy  
\begin{eqnarray}
{\bf u}_L ({\bf x},t) - {\bf u}_E ({\bf x},t)\propto \nabla \ln n ({\bf x},t) \, .
\end{eqnarray}
This relation is analogous of the consistency condition (\ref{eqn:cc2}).
In addition, if we require the positivity of the kinetic term of the stochastic Lagrangian, we cannot set $\kappa=0$ and thus the contribution of the $\Pi^{ij}_Q ({\bf x},t)$ term should be maintained.
See Appendix \ref{app:positivity}.

As was mentioned in the last of Sec.\ \ref{sec:idealfluid_vari}, we need to assume no intersection of the trajectories of different fluid elements to show the equivalence of the field-type action (\ref{eqn:idealaction1}) and the particle-type action (\ref{eqn:idealaction2}). Once the fluctuations of the trajectories are introduced, however, this will not be satisfied and thus we cannot find the corresponding field-type action from Eq.\ (\ref{eqn:sto_ac_fluid}).
The applications of SVM to field-theoretical systems have not yet been studied sufficiently except for the Klein-Gordon field  \cite{koide_field} and thus it is still an open question whether the NSF equation is obtained from the field-type action (\ref{eqn:ideal_eular_lag}) by introducing zigzag fields.

\subsection{Uncertainty relations in hydrodynamics}

In the discussion of the uncertainty relations in quantum mechanics, 
we consider the position and the momentum of a quantum particle. 
For the uncertainty relations in hydrodynamics, we consider those of fluid elements.
Suppose that each fluid element has a fixed common mass $\uM$ which is much smaller than the total mass of the fluid, 
$\uM << \uM_T$ to justify the local thermal equilibrium.
Then the total number of fluid elements is 
given by 
\begin{eqnarray}
N_{eff} = \frac{\uM_T}{\uM} = \int \ud^D x \, \rho^{eff} ({\bf x},t) \, , \label{eqn:mtm}
\end{eqnarray}
where we introduced the distribution of fluid elements,  
\begin{eqnarray}
\rho^{eff} ({\bf x},t) = \frac{1}{\uM} n({\bf x},t) \, .
\end{eqnarray}
The choice of $\uM$ and hence the definition of $\rho^{eff} ({\bf x},t)$ depends on our definition 
of fluid elements. 
As is seen soon later, however, the minimum value of uncertainty is independent of the ambiguity of the definition.

Then the standard deviation of the positions of fluid elements will be defined by 
\begin{eqnarray}
\left( \sigma^{(2)}_{x^i} \right)^{1/2} = \sqrt{\lceil (\delta \widehat{r}^i )^2 \rfloor} \, .
\end{eqnarray}
The definition of $\delta$ is the same as that given below Eq.\ (\ref{eqn:sigma-x-particle}), 
$\delta \widehat{r}^i = \widehat{r}^i ({\bf R},t)- \lceil \widehat{r}^i \rfloor$, where, however, 
the definition of  the average $\lceil~~~\rfloor$ is replaced by 
\begin{eqnarray}
\lceil \widehat{f} \, \rfloor 
= \int \ud^D R \,  \rho^{eff}_{0} ({\bf R}) \uE[f(\widehat{\bf r}({\bf R},t),t)] 
= \int \ud^D x \,  \rho^{eff} ({\bf x},t) f({\bf x},t)
\, . \label{eqn:new-ave}
\end{eqnarray}
Here we used the initial distribution of fluid elements $\rho^{eff}_{0} ({\bf x}) = \rho^{eff} ({\bf x},t_i)$. 

The standard deviation of the momentum of fluid elements 
is defined in the same fashion as Eq.\ (\ref{delta_p}) using Eq.\ (\ref{eqn:new-ave}), 
\begin{eqnarray}
\left( \sigma^{(2)}_{p^{i}} \right)^{1/2} =\sqrt{ \frac{\lceil (\delta \widehat{p}^{i}_+ )^2 \rfloor
+ \lceil (\delta \widehat{p}^{i}_-  )^2 \rfloor  }{2} } \, ,
\end{eqnarray}
where the two momenta are 
\begin{eqnarray}
\left(
\begin{array}{c}
\widehat{\bf p}_+ (t) \\ 
\widehat{\bf p}_- (t) 
\end{array}
\right)
=
2\uM
\left(
\begin{array}{c}
A_+ B_+ {\bf u}_+ ({\bf r}({\bf R},t),t) + \frac{1}{2} B_- {\bf u}_- ({\bf r}({\bf R},t),t)
\\
\frac{1}{2} B_- {\bf u}_+ ({\bf r}({\bf R},t),t) + A_- B_+ {\bf u}_- ({\bf r}({\bf R},t),t)
\end{array}
\right) \, . \label{eqn:fluid_mom}
\end{eqnarray}
Using these, we find 
\begin{equation}
\begin{split}
\lceil \widehat{r}^{i} \widehat{p}^{j}_{\um}\, \rfloor
&=
- 2\uM_T \nu \alpha_A B_+ \delta_{ij} + \int \ud^D x \, n ({\bf x},t) x^{i} v^{j} ({\bf x},t) \, ,\\
\lceil \widehat{r}^{i} \widehat{p}^{j}_{\ud}\, \rfloor
&=
- 2\uM_T \nu \alpha_B \delta_{ij} + 2 B_+ \alpha_A  \int \ud^D x \, n ({\bf x},t) x^{i} v^{j} ({\bf x},t) 
\, ,
\end{split}
\label{eqn:xpdm-fluid}
\end{equation}
where the total mass of the fluid $\uM_T$ comes from Eq.\ (\ref{eqn:mtm}).

The uncertainty relation in hydrodynamics is eventually given by
\begin{eqnarray}
\left( \sigma^{(2)}_{x^{i}}\right)^{1/2} \left( \sigma^{(2)}_{p^{j}} \right)^{1/2}  
\ge 
\uM_T \sqrt{
\frac{(\xi^2-\kappa)^2}{\nu^2 + \xi^2} \delta_{ij} 
+ \left( 1 + \frac{\xi^2}{\nu^2} \right) 
\left(
\frac{\uM}{\uM_T}\lceil \delta \widehat{r}^{i}  \, \delta \widehat{v}^{j}  \rfloor
- 
\frac{\xi (\nu^2 + \kappa)}{\nu^2 + \xi^2}
\delta_{ij}
\right)^2 }\, . \nonumber \\
\end{eqnarray}
Here we introduced the kinematic viscosity 
\begin{eqnarray}
\xi = \frac{\eta}{2n} = 2 \alpha_A B_+  \nu \, . \label{eqn:kv}
\end{eqnarray}
This is the Robertson-Schr\"{o}dinger-type inequality in hydrodynamics. 
The right-hand side is represented by the parameters in hydrodynamics showing that 
the magnitude of the minimum value is characterized by the viscosity, which 
is known to be associated with thermal fluctuations through the fluctuation-dissipation theorem.

The second term inside the square root on the right-hand side depends on the behaviors of fluids 
but is always positive. 
Therefore, by dropping this, the above inequality is simplified 
and we obtain the Kennard-type inequality in hydrodynamics, 
\begin{eqnarray}
\left( \sigma^{(2)}_{x^{i}} \right)^{1/2} \left( \sigma^{(2)}_{p^{j}} \right)^{1/2}  
\ge 
\uM_T  
\frac{|\xi^2-\kappa|}{\sqrt{\nu^2 + \xi^2}} \delta_{ij} 
=
\frac{4\uM_T \nu}{\sqrt{1+ (\xi/\nu)^2}}|{\rm det} ({\cal M})| \delta_{ij} 
\, , \label{eqn:fluid_uc}
\end{eqnarray}
where
\begin{eqnarray}
{\rm det} ({\cal M}) = \frac{\kappa - \xi^2}{4\nu^2} \neq 0 \, .
\end{eqnarray}
It should be emphasized that 
the minimum value of uncertainty is independent of the definition of fluid elements.
This depends on the total mass of the fluid $\uM_T$ and thus becomes larger as 
the increase of  the volume of fluids.

In the above definition of $\sigma_{x^i}$, we measure the position from 
the following expectation value,
\begin{eqnarray}
\lceil \widehat{r}^i \, \rfloor = \int \ud^D x \, \rho^{eff} ({\bf x},t) x^i \, .
\end{eqnarray}
On the other hand, the center of mass of fluids is defined by 
\begin{eqnarray}
\frac{1}{\int \ud^D x\, \rho^{eff}({\bf x},t) } \lceil \widehat{r}^i \, \rfloor  \, ,
\end{eqnarray}
and then we can define the standard deviation of the position measured from the center of mass by 
\begin{eqnarray}
\left( \Sigma^{(2)}_{x^{i}} \right)^{1/2}  = \frac{1}{\int \ud^D x \, \rho^{eff}({\bf x},t)} \left( \sigma^{(2)}_{x^i} \right)^{1/2} \, .
\end{eqnarray}
Using this definition, the Kennard-type inequality is reexpressed as 
\begin{eqnarray}
\left( \Sigma^{(2)}_{x^{i}} \right)^{1/2} \left( \sigma^{(2)}_{p^{j}} \right)^{1/2}  
\ge 
 \frac{4\uM \nu}{\sqrt{1+ (\xi/\nu)^2}}|{\rm det} ({\cal M})| \delta_{ij} 
\, .
\end{eqnarray}
Here $\uM_T$ is replaced with $\uM$ and thus this minimum value depends on the 
choice of the definition of the fluid element.
This mass cannot be smaller than the mass of 
the constituent particle of a simple fluid, $m_{cs}$. 
Thus, 
by substituting $m_{cs}$ into $\uM$, 
the Kennard-type inequality should satisfy
\begin{eqnarray}
\left( \Sigma^{(2)}_{x^{i}} \right)^{1/2}  \left( \sigma^{(2)}_{p^{j}} \right)^{1/2}  
\ge 
 \frac{4 m_{cs} \nu}{\sqrt{1+ (\xi/\nu)^2}}|{\rm det} ({\cal M})| \delta_{ij} 
\, .
\end{eqnarray}

To estimate the right-hand side, 
we consider water at room temperature, 
where the mass of the molecule is $\sim 3 \times 10^{-26}$ kg  and 
$\xi \sim 10^{-6} \, {\rm m}^2/s$.
The stochastic intensity $\nu$ may be identified with a diffusion coefficient in fluid which is normally much smaller than 
$\xi$, say, $\nu \sim 10^{-9}$ \, ${\rm m}^2/s$.
Then the minimum value for water becomes 
\begin{eqnarray}
 \frac{4 m_{cs} \nu}{\sqrt{1+ (\xi/\nu)^2}}|{\rm det} ({\cal M})| \sim m_{cs} \xi 
\sim 3 \times 10^{-32} \, [{\rm kg \, m^2/s}] \sim 600 \times \frac{\hbar}{2}
\, . \label{eqn:water_uc}
\end{eqnarray}
On the other hand, we choose $\xi \sim 0.3 \times 10^{-6}$ \, ${\rm m}^2/s$ and  $\nu \sim 10^{-4}$ \, ${\rm m}^2/s$ for water vapor and then the minimum value becomes
\begin{eqnarray}
 \frac{4 m_{cs} \nu}{\sqrt{1+ (\xi/\nu)^2}}|{\rm det} ({\cal M})| \sim 60 \times \frac{\hbar}{2}
\, .
\end{eqnarray}
In these calculations, we set $\kappa = 0$ to consider the NSF equation.
These values are larger than that of quantum mechanics by two or three order
but will be much smaller than the typical coarse-grained scales of
hydrodynamics. 
To observe the minimum value of uncertainty, we need to measure the mass
distribution and the velocity field with precision. In such a
precision, however, the hydrodynamical description normally looses its
validity because the hydrodynamical approach is applicable only to macroscopic
variables observed in a coarse-grained scale.

We have emphasized that the fluid element should be chosen so as to satisfy the local thermal equilibrium.
Mathematically, this condition affects only the definition of the potential term of the Lagrangian. 
On the other hand,  
the uncertainty relations in our formulation are derived from the consistency condition (\ref{eqn:cc}) and the definitions of momenta (\ref{eqn:lad-tra}), and thus independent of the property of the potential term. 
Therefore, our relations are applicable to continuum media where the local thermal equilibrium is not satisfied.

\subsection{Uncertainty relations in quantum field theory and SVM} \label{sec:qft_uc}

The Gross-Pitaevskii equation describes a coarse-grained quantum dynamics of, for example, the Bose-Einstein condensate (BEC) in the trapped Bose gas. 
This equation is normally derived by applying the mean-field approximation to the bosonic Schr\"{o}dinger-field equation.
Let us consider the uncertainty relation for such a field-theoretical system.

The bosonic Schr\"{o}dinger-field operator satisfies the canonical commutation rule,  
\begin{eqnarray}
[\phi({\bf x},t), \phi^\dagger ({\bf x},t)] = \delta^{(D)} ({\bf x} -{\bf x}^\prime) \, .
\end{eqnarray} 
Using these field operators, the field-theoretical position and momentum operators are defined by 
\begin{eqnarray}
x^i_{op} &=& \int \ud^D x\, \phi^\dagger ({\bf x},t) x^i \phi({\bf x},t) \, ,\\
p^j_{op} &=& \int \ud^D x\, \phi^\dagger ({\bf x},t) (-\ii \hbar \partial_{j}) \phi({\bf x},t) \, ,
\end{eqnarray}
respectively. 
One can easily confirm that these operators satisfy
\begin{eqnarray}
[x^i_{op}, \, p^j_{op}] = \ii \hbar N_{op} \delta_{ij} \, ,
\end{eqnarray}
where the number operator is defined by
\begin{eqnarray}
N_{op} = \int \ud^D x\, \phi^\dagger ({\bf x},t) \phi({\bf x},t) \, .
\end{eqnarray}

Then the Kennard inequality for this many-body system is calculated as
\begin{eqnarray}
\sigma_{x^i} \sigma_{p^j} \ge \frac{\hbar}{2} \langle N_{op} \rangle \delta_{ij} \, ,\label{eqn:qft_uc}
\end{eqnarray} 
where $\langle ~~~~ \rangle$ denotes the standard expectation value with a Fock state vector 
in quantum field theory. Then the above standard deviations are defined by 
\begin{eqnarray}
\sigma_A = \sqrt{ \langle (A_{op} -\langle A_{op} \rangle)^2 \rangle} \, ,   \label{eqn:qm_sd} 
\end{eqnarray}
where $A_{op}$ is an operator.

This uncertainty relation is reproduced in SVM \cite{koide18}. 
As is discussed in Ref.\ \cite{koide18} and summarized in Appendix \ref{app:gp}, 
the Gross-Pitaevskii equation is obtained by applying 
the SVM quantization to the classical Lagrangian of the ideal fluid (\ref{eqn:idealaction2}).
To obtain the corresponding uncertainty relation from Eq.\ (\ref{eqn:fluid_uc}), we set $(\alpha_A,\alpha_B,\nu) = (0,1/2,\hbar/(2\uM))$,
\begin{eqnarray}
\left( \sigma^{(2)}_{x^{i}} \right)^{1/2} \left( \sigma^{(2)}_{p^{j}} \right)^{1/2}  
\ge 
\frac{\hbar}{2} N \delta_{ij} \, . 
\end{eqnarray}
Note that $\uM$ is the mass of the constituent particle of BEC and $N = \uM_T/\uM = \langle N_{op} \rangle$ is the number of particles. This coincides with the result from quantum field theory, (\ref{eqn:qft_uc}).

\section{Numerical examples and uncertainty relations for liquid and gas} \label{sec:numerical}

To investigate the behavior of the uncertainty relations in hydrodynamics, 
we consider the time evolution of the fluid described by the NSF equation in 1+1 dimensional system.  
As the initial condition, we choose the static fluid ($v (x,0) = 0$) 
with the Gaussian mass distribution 
\begin{eqnarray}
n (x,0) = \frac{\uM_T}{\sqrt{2\pi}x_0} e^{- \frac{x^2}{2x^2_0}} 
\equiv \frac{n_0}{\sqrt{2\pi}} e^{- \frac{x^2}{2x^2_0}} \, ,
\label{eqn:gidis}
\end{eqnarray}
where $x_0$ is a parameter characterizing the spatial distribution.

Moreover, we are interested in the behavior of the NSF equation and hence we 
set $\alpha_B =0$ (or equivalently $\kappa=0$) in the following numerical calculations.

\subsection{Time evolution of the NSF equation in 1+1 dimension}

\begin{figure}[t]
\begin{center}
\includegraphics[scale=0.3]{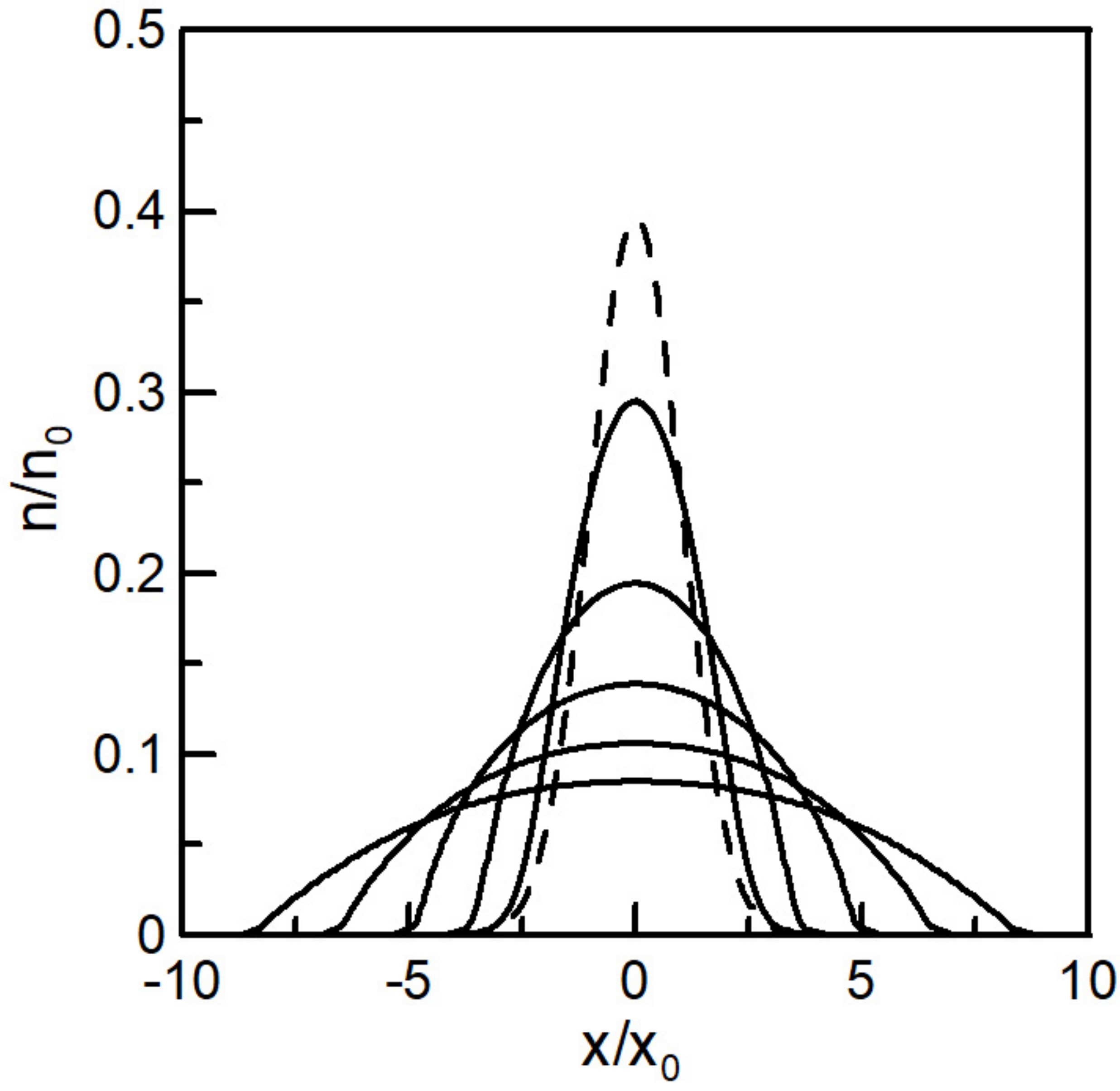}
\includegraphics[scale=0.3]{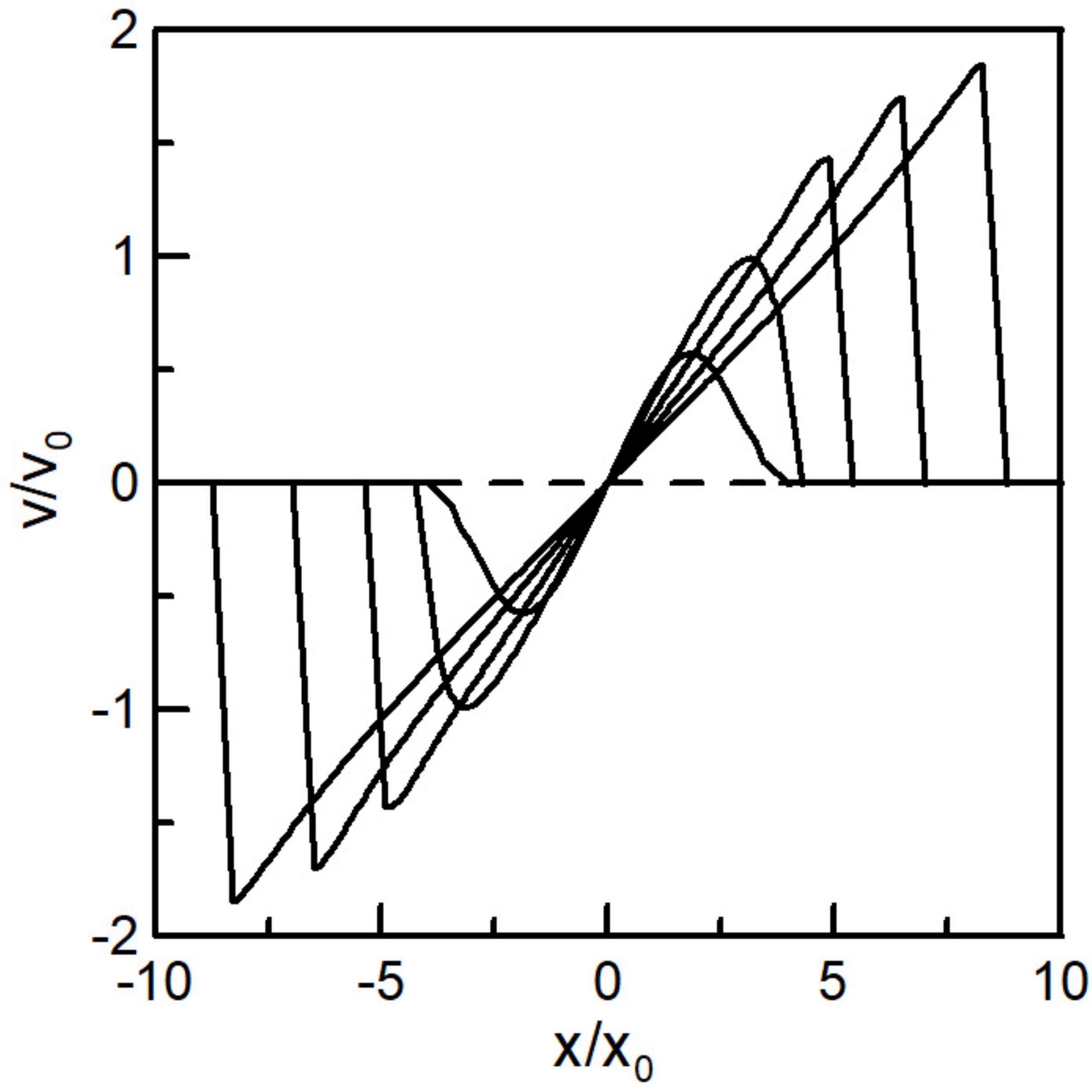}
\end{center}
\caption{
The time evolutions of the mass distribution and the velocity field 
are shown on the left and right panels, respectively.
The Reynolds number is set by $\RE=10$.
The initial values are denoted by the dashed lines.
The five different solid lines represent the results for $\tau = 1,2,3,4$ and $5$, respectively 
}
\label{fig:nsf}
\end{figure}

The adimensional representation of the NSF equation has a unique parameter, 
the Reynolds number ($\RE$), defined by 
\begin{eqnarray}
\RE = \frac{x_0 v_0}{\zeta}n = \frac{x_0 v_0}{\eta}n  \, ,
\end{eqnarray}
where $v_0$ is a typical scale of the fluid velocity.

Then the NSF equation in the adimensional form is represented by
\begin{eqnarray}
(\partial_\tau + \underline{v} \partial_q ) \underline{v} = -\frac{1}{\underline{n}} 
\partial_q \left( 
\underline{P}  - \frac{\underline{n}}{\RE} \partial_q \underline{v} 
\right) \, ,
\end{eqnarray} 
where adimensional quantities are introduced as 
\begin{equation}
\begin{split}
q &= \frac{x}{x_0} \, ,\\
\underline{v} &= \frac{v}{v_0} \, ,\\
\tau &= t \frac{v_0}{x_0} \equiv \frac{t}{t_0} \, ,\\
\underline{n} &= \frac{n}{n_0} \, , \\
\underline{P} &= P\frac{1}{n_0 v^2_0} \, .
\end{split}
\end{equation}
We further consider the pressure in the adiabatic (isentropic) process of the ideal gas, 
\begin{eqnarray}
\underline{P} = C \, \underline{n}^{5/3} \, .
\end{eqnarray}
For the sake of simplicity, we set the coefficient $C = 1$.

The numerical algorithm to solve the NSF equation is smoothed particle hydrodynamics (SPH). For the brief summary of SPH, see Appendix \ref{app:sph}. 
The mass distribution and the velocity field of the fluid are calculated and shown on the left and right panels 
of Fig.\ \ref{fig:nsf}, respectively. 
The Reynolds number is set by $\RE=10$. The initial values, which are given 
by the static Gaussian distribution mentioned above, are denoted by the dashed lines. 
The five different solid lines represent the results for $\tau = 1,2,3,4$ and $5$, respectively

\begin{figure}[t]
\begin{center}
\includegraphics[scale=0.3]{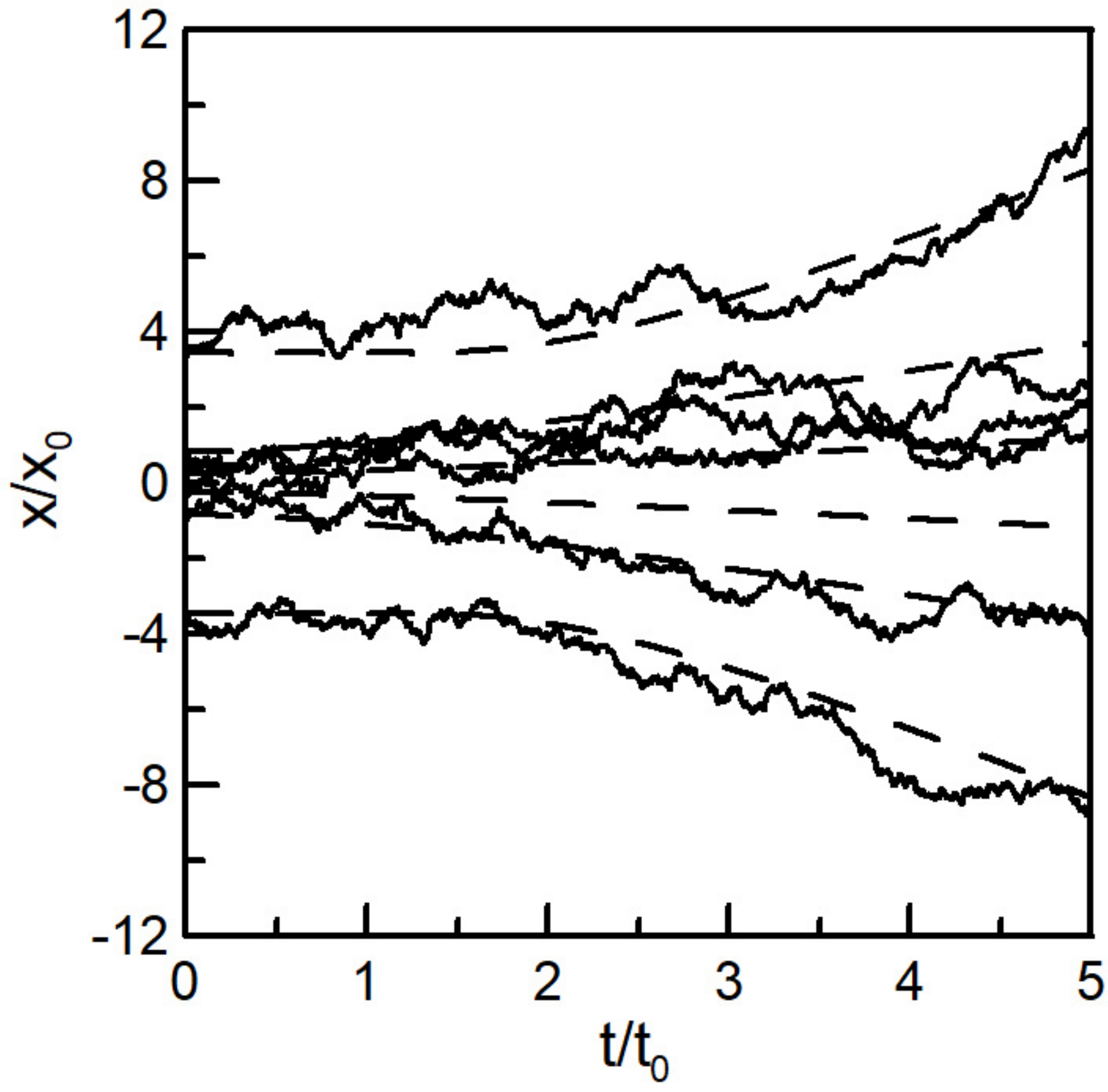}
\includegraphics[scale=0.3]{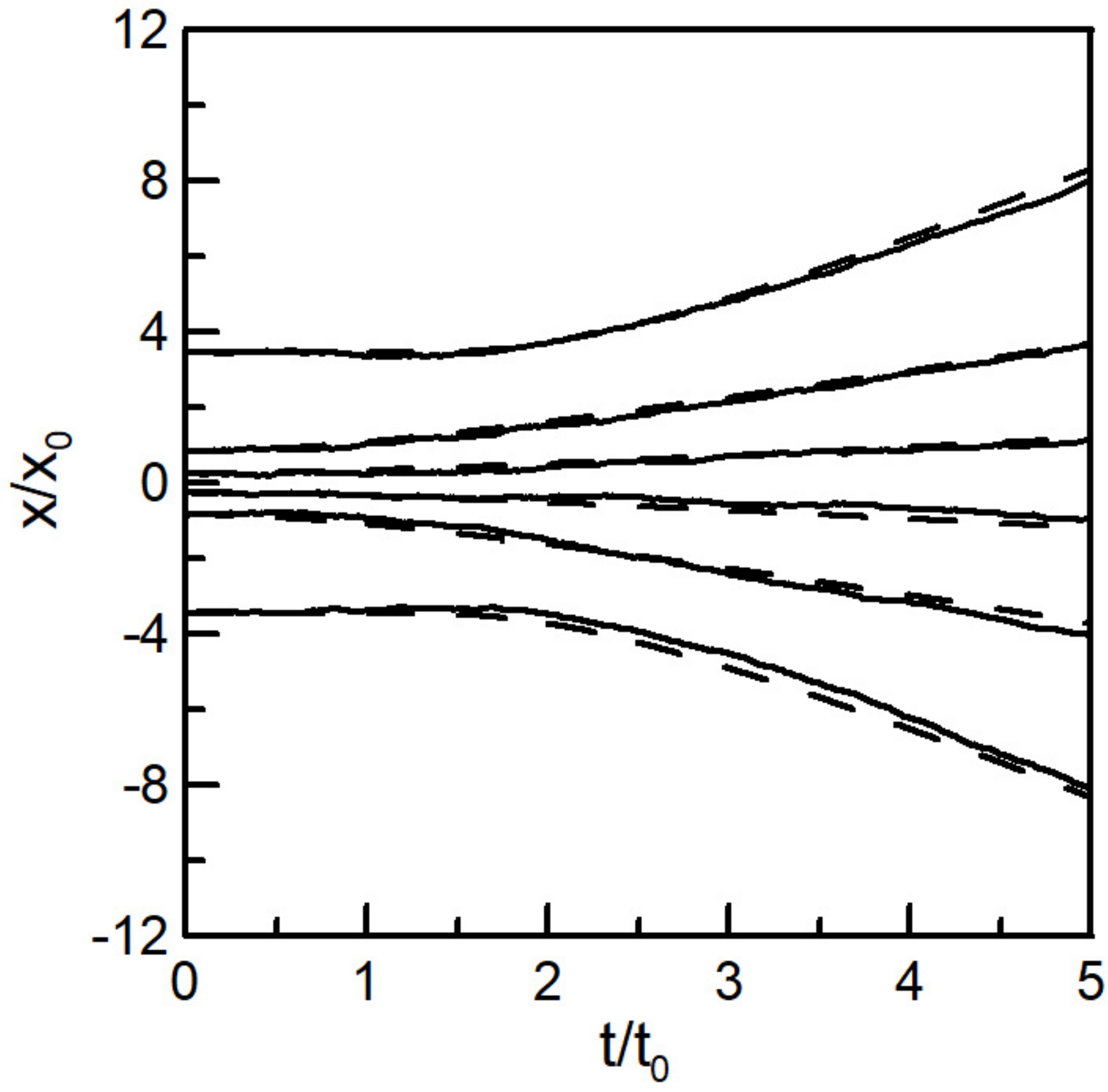}
\end{center}
\caption{
The trajectories of Brownian motions of fluid elements. 
The left and right panels represent the results for $\alpha_A=0.1$ and $10$, respectively.
The six different initial positions are chosen by $q(0) =-3.45,-0.84,-0.25,0.25,0.84$ and $3.45$.
For the sake of comparison, the streamlines of the fluid are shown by the dotted lines.  
}
\label{fig:bm_fluidelement}
\end{figure}

Brownian motions of fluid elements are calculated from the forward SDE (\ref{eqn:fsde}) by substituting $n(x,t)$ 
and $v(x,t)$ of the NSF equation.
It should be noted that the fluctuations of the trajectories depend on another parameter $\alpha_A$ 
which does not appear explicitly in the NSF equation. 
In Fig.\ \ref{fig:bm_fluidelement}, we show the trajectories of the fluid elements for two different values of $\alpha_A$ 
fixing $\RE=10$. The left panel is the result for $\alpha_A=0.1$ and the right panel for $\alpha_A=10$. 
The forward SDE (\ref{eqn:fsde}) is numerically solved using the Euler-Maruyama method with $\ud \tau =0.005$. 
The six different initial positions of the fluid elements are utilized to show the stochastic trajectories; $q(0)=-3.45,-0.84,-0.25,0.25,0.84$ and $3.45$.
For the sake of comparison, the streamlines of the fluid are shown by the dotted lines.  
One can see that the fluctuation of the trajectory is enhanced for the smaller value of $\alpha_A$ because the intensity of the thermal noise $\nu$ can be expressed as 
\begin{eqnarray}
\nu = \frac{x_0 v_0}{2\alpha_A \RE} \, .
\end{eqnarray}

\subsection{uncertainty relations for Gaussian initial condition} \label{sec:nc_uc}

The adimensional representation of the standard deviation of position  is given by
\begin{eqnarray}
\underline{\sigma}^{(2)}_{x^{i}}(\tau) = \int \ud^D q \, \underline{n} ({\bf q},\tau) \left( 
q^i - \frac{\uM_T}{\uM} \int \ud^D q \, \underline{n}({\bf q},\tau) q^i 
\right)^2 
= \frac{\uM}{\uM_T x^2_0}\sigma^{(2)}_{x^{i}} \, .
\end{eqnarray}
It should be noted that the two masses, $\uM$ and $\uM_T$, appear because the definition of $\rho^{eff}({\bf q},\tau)$ depends on the choice of the mass of the fluid element. 
For the corresponding quantity for momentum, we define  
\begin{eqnarray}
\underline{\sigma}^{(2)}_{p^i} (\tau)
= \frac{\uM^2_T}{\uM^2} \int \ud^D q \, \underline{n} ({\bf q},\tau) d^2_{p^i} ({\bf q},\tau) 
= \frac{1}{\uM \uM_T v^2_0} \sigma^{(2)}_{p^i} \, . \label{eqn:flu-mom-sd}
\end{eqnarray}
Here an adimensional function is introduced by 
\begin{eqnarray}
d^2_{p^i} ({\bf q},\tau) = \frac{1}{2} \sum_{a=\pm} \left[
(\varpi^i_a ({\bf q},\tau))^2 - \left( \frac{\uM_T}{\uM} \int \ud^D q \, \underline{n} ({\bf q},\tau) \varpi^i_a ({\bf q},\tau) \right)^2
\right] \, ,
\end{eqnarray}
with 
\begin{eqnarray}
\left(
\begin{array}{c}
\varpi^i_+ ({\bf q},\tau) \\ 
\varpi^i_- ({\bf q},\tau) 
\end{array}
\right)
&=& 
\frac{\uM}{\uM_T v_0}
\left(
\begin{array}{c}
A_+  {u}^i_+ ({\bf x},t) + \frac{1}{2}  {u}^i_- ({\bf x},t)
\\
\frac{1}{2} {u}^i_+ ({\bf x},t) + A_-  {u}^i_- ({\bf x},t)
\end{array}
\right)\, .
\end{eqnarray}
These $\varpi^i_\pm ({\bf q},\tau)$ are adimensional representations of the momentum functions $p_{\pm} ({\bf x},t)$ which are obtained from Eq.\ (\ref{eqn:fluid_mom}) as
\begin{eqnarray}
\left(
\begin{array}{c}
{ p}_+ ({\bf x},t) \\ 
{ p}_- ({\bf x},t) 
\end{array}
\right)
=
2\uM
\left(
\begin{array}{c}
A_+ B_+ { u}_+ ({\bf x},,t) + \frac{1}{2} B_- { u}_- ({\bf x},t)
\\
\frac{1}{2} B_- { u}_+ ({\bf x},t) + A_- B_+ {u}_- ({\bf x},t)
\end{array}
\right) \, , \label{eqn:momfunc}
\end{eqnarray}
with $\alpha_B = 0$.

Then the Kennard-type uncertainty relation in terms of the adimensional quantities is 
expressed as 
\begin{eqnarray}
\left( \underline{\sigma}^{(2)}_{x^i} (\tau) \right)^{1/2}
\left( \underline{\sigma}^{(2)}_{p^j} (\tau) \right)^{1/2}
\ge 
\frac{1}{\RE} \sqrt{\frac{\alpha^2_A}{4(1+\alpha^2_A)}} \delta_{ij} \, . 
\end{eqnarray}
For the symmetric initial condition considered here, 
$\underline{\sigma}^{(2)}_{x^i} (\tau)$ and $\underline{\sigma}^{(2)}_{p^j} (\tau)$ 
are independent of $\uM$. 
Thus our simulations do not have ambiguity associated with the choice of the mass of the fluid element.

Using the Gaussian initial distribution defined by Eq.\ (\ref{eqn:gidis}), the initial value of the product is estimated exactly as
\begin{eqnarray}
( \underline{\sigma}^{(2)}_{x} (0) )^{1/2} ( \underline{\sigma}^{(2)}_{p} (0) )^{1/2} = \frac{1}{2\RE} 
> \frac{1}{\RE} \sqrt{\frac{\alpha^2_A}{4(1+\alpha^2_A)}}
\, . \label{eqn:ini_uc}
\end{eqnarray}
The initial value is always larger than the minimum value for any real $\alpha_A$.

As was discussed in Sec.\ \ref{sec:cnsf}, 
the parameter $\alpha_A$ is given by the ratio of the kinematic viscosity $\xi$ and the diffusion coefficient $\nu$: 
$\alpha_A >> 1$ for liquid and $\alpha_A >> 1$ for gas in water.
Thus we investigate the uncertainty relations for two different values, $\alpha_A = 10$ and $\alpha_A = 0.1$, 
and, for the sake of convenience, we call the former liquid case and the latter gas case, respectively.

\begin{figure}[t]
\begin{center}
\includegraphics[scale=0.3]{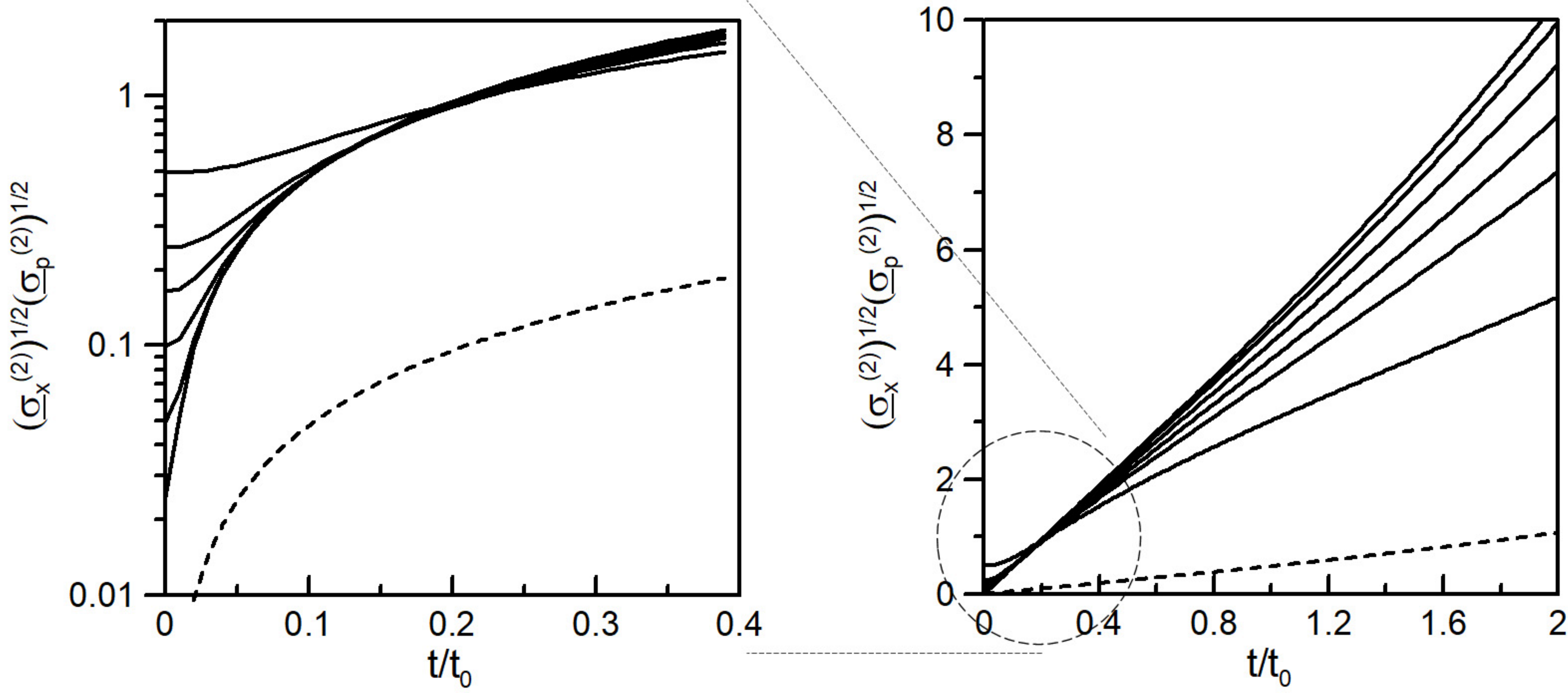}
\end{center}
\caption{
The time evolutions of the product $(\underline{\sigma}^{(2)}_{x} (\tau))^{1/2} (\underline{\sigma}^{(2)}_{p} (\tau) )^{1/2}$ for the liquid case ($\alpha_A = 10$) are shown on the right panel. 
The extended figure for the early stage of the time evolution is plotted on the left panel in logarithmic scale. 
The six solid lines represent the results for $\RE=1$, $2$, $3$, $5$, $10$ and $20$. 
For the sake of comparison, the result for the ideal fluid is denoted by the dotted line. 
The inclination of the solid line is enhanced as $\RE$ is increased.
All solid lines are always larger than the theoretically predicted minimum values.
}
\label{fig:uc_evol_alp=10}
\end{figure}

\begin{figure}[t]
\begin{center}
\includegraphics[scale=0.3]{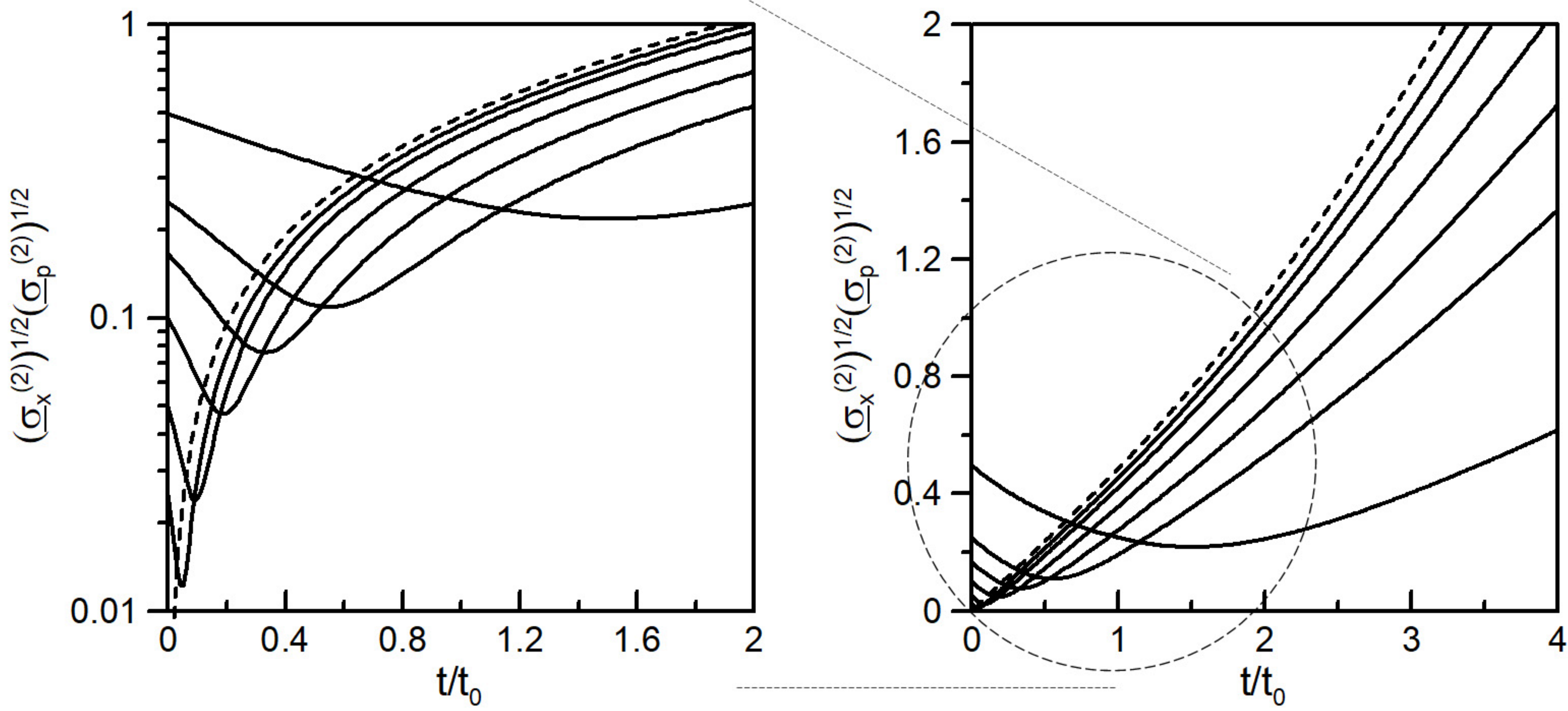}
\end{center}
\caption{
The time evolutions of the product $(\underline{\sigma}^{(2)}_{x} (\tau))^{1/2} (\underline{\sigma}^{(2)}_{p} (\tau) )^{1/2}$ for the gas case ($\alpha_A = 0.1$) are shown on the left panel. 
The extended figure for the early stage of the time evolution is plotted on the right panel in logarithmic scale. 
The six solid curves represent the results for $\RE=1$, $2$, $3$, $5$, $10$ and $20$.
For the sake of comparison, the result for the ideal fluid is denoted by the dotted line. 
The position of the local minimum of the solid curve moves to the upper right as $\RE$ is decreased.
All solid curves are always larger than the theoretically predicted minimum values.
}
\label{fig:uc_evol_alp=0_1}
\end{figure}

The time evolution of the product $(\underline{\sigma}^{(2)}_{x} (\tau))^{1/2}(\underline{\sigma}^{(2)}_{p} (\tau))^{1/2}$ 
for the liquid case ($\alpha_A = 10$) is shown on the right panel of Fig.\ \ref{fig:uc_evol_alp=10}.
The extended figure for the early stage of the time evolution is plotted on the left panel in logarithmic scale. 
The six solid lines represent the results for $\RE=1$, $2$, $3$, $5$, $10$ and $20$. 
For the sake of comparison, the dotted line denotes the 
result for the ideal fluid, which is calculated by setting $(\alpha_A,\alpha_B, \nu)= (0,0,0)$ in the definition of the standard deviations and substituting the numerical result of the Euler equation.
The inclination of the solid line is enhanced as $\RE$ is increased.
All solid lines are monotonically increasing functions and always stay above the result of the ideal fluid.
Because of Eq.\ (\ref{eqn:ini_uc}), then, it is easy to see that these lines are always larger than the theoretically predicted minimum values. 
Except for the early stage of the time evolution, the line for smaller $\RE$ is closer to the ideal-fluid line.

The behavior of the product $(\underline{\sigma}^{(2)}_{x} (\tau))^{1/2}(\underline{\sigma}^{(2)}_{p} (\tau))^{1/2}$ for the gas case ($\alpha_A = 0.1$) is qualitatively different as is shown on the right panel of  Fig.\ \ref{fig:uc_evol_alp=0_1}.
The extended figure for the early stage of the time evolution is plotted on the left panel in logarithmic scale. 
The six solid curves represent the results for $\RE=1$, $2$, $3$, $5$, $10$ and $20$.
The result for the ideal fluid is denoted by the dotted line. 
The position of the local minima of the solid curves moves to the upper right as $\RE$ is decreased.
Because of the initial condition (\ref{eqn:ini_uc}), all curves start from the larger values than that of the ideal fluid. 
The solid lines behave as decreasing functions in the early stage and, afterward, become increasing functions which stay below the result of the ideal fluid.
Although it is not explicitly shown, these local minima of the solid lines are always larger than the 
theoretically predicted minimum values.

In these figures, 
the solid line of larger $\RE$ stays above that of smaller $\RE$ 
in the late stage of the time evolution.
It is because both of the spreading of the mass of the fluid and the evolution of the velocity field 
are decelerated as $\RE$ decreases due to the enhancement of the effect of viscosity.

In these simulations, we find that the product for the liquid case ($\alpha_A =10$) is always larger 
than that of the gas case ($\alpha_A =0.1$). 
This can be understood from the definition of the fluid momenta.
As is seen from Eq.\ (\ref{eqn:momfunc}), two momenta are given by the linear combination of the two contributions: 
one is the fluid velocity ${\bf v}({\bf x},t)$ and the other $\nabla \ln n ({\bf x},t)$. In our simulations, 
these are given by   
\begin{equation}
\begin{split}
p_+ (x,t) &= \uM \left\{ (1+\alpha_A) v (x,t) + \frac{x_0 v_0}{2\RE} \partial_x \ln n (x,t) \right\} \, ,\\ 
p_- (x,t) &= \uM \left\{ (1-\alpha_A) v (x,t) + \frac{x_0 v_0}{2\RE} \partial_x \ln n (x,t) \right\} \, .
\end{split}
\end{equation}
We focus on the behavior at $x \ge 0$ in our Gaussian initial condition. Then as is seen from the left panel of Fig.\ \ref{fig:nsf}, 
the contribution of $\partial_x \ln n (x,t)$ is negative but that of $v(x,t)$ is positive. 
Thus a part of the contributions from the two terms is canceled for any real $\alpha_A \ge 0$ in $p_+ (x,t)$. 
The same discussion is applied to $p_- (x,t)$ when $0 \le \alpha_A \le 1$, but  
such a cancellation does not occur for $\alpha_A >1$.
Therefore the standard deviation of momentum for liquid ($\alpha_A = 10$) is always larger than that of gas ($\alpha_A = 0.1$). The same argument is applicable to the behavior at $x<0$. 
This may suggest that the qualitatively different behavior in the uncertainty relation may be used to classify liquid and gas in fluids.

\begin{figure}[t]
\begin{center}
\includegraphics[scale=0.3]{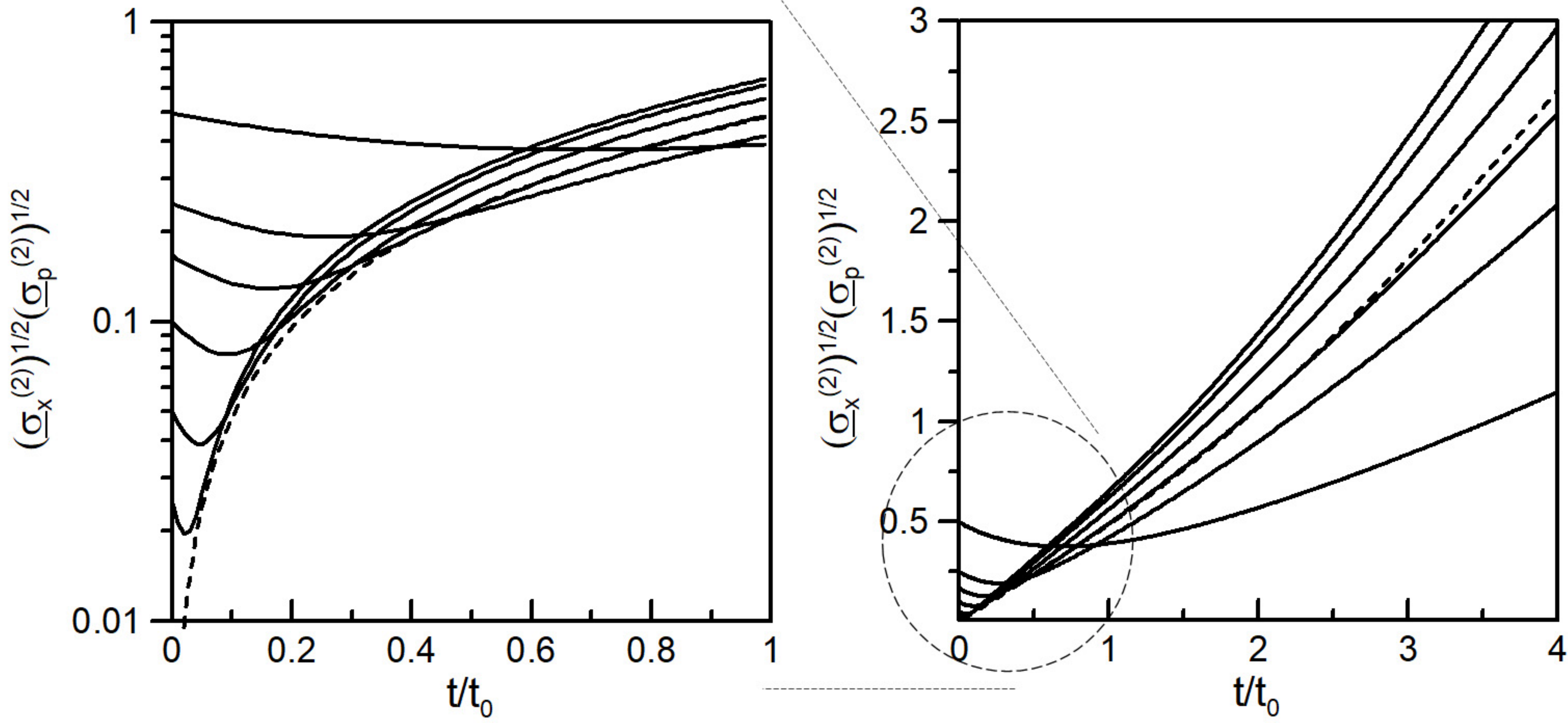}
\end{center}
\caption{
The time evolutions of the product $(\underline{\sigma}^{(2)}_{x} (\tau))^{1/2} (\underline{\sigma}^{(2)}_{p} (\tau) )^{1/2}$ for the fluid at the boundary between liquid and gas ($\alpha_A = 1$) are shown on the right panel. 
The extended figure for the early stage of the time evolution is plotted on the left panel in logarithmic scale. 
The six solid curves represent the results for $\RE=1$, $2$, $3$, $5$, $10$ and $20$.
For the sake of comparison, the result for the ideal fluid is denoted by the dotted line. 
The position of the local minimum of the solid curves moves to the upper right as $\RE$ is decreased.
The solid line of $\RE=3$ and the dotted line are almost on top of each other.
The solid curves for $\RE \ge 5$ stay above the dotted line.
}
\label{fig:uc_evol_alp=1}
\end{figure}

To see the behavior of the fluid at the boundary of the liquid and gas cases, 
the products for $\alpha_A = 1$ are shown 
on the right panel of Fig.\ \ref{fig:uc_evol_alp=1}. 
The extended figure for the early stage of the time evolution is plotted on the left panel in logarithmic scale. 
The six solid curves represent the results for $\RE=1$, $2$, $3$, $5$, $10$ and $20$.
The result for the ideal fluid is denoted by the dotted line. 
The position of the local minima of the solid curves moves to the upper right as $\RE$ is decreased.
Differently from the previous calculations, the asymptotic behavior depends on $\RE$: 
the solid curves for $\RE \ge 5$ stay above the dotted line but those for $\RE \le 2$ stay below.
The solid line of $\RE=3$ and the dotted line are almost on top of each other.

Suppose that the dotted line for the ideal fluid gives a dividing ridge of liquid and gas. 
For the previous simulations for liquid and gas, 
the solid lines never intersect the dotted line in the late stage of the time evolution.
The intersects in the early stage are affected by the choice of initial conditions and 
we focus only on the asymptotic behaviors of the uncertainty relation.
Then the classification of liquid and gas does not depends on the values of $\RE$ for $\alpha_A = 0.1$ and $10$.
On the other hand, the fluid of $\alpha_A=1$ changes its behavior depending on $\RE$.
There is an critical value $\RE^*$ near $\RE=3$. 
For the fluid with $\RE> \RE^*$, the asymptotic behavior of the product is larger than the ideal one and thus 
is classified as liquid. 
On the other hand, the fluid with $\RE < \RE^*$ is classified as gas.

\section{Discussions and concluding remarks}  \label{sec:conclusion}

We showed that the quantum-mechanical uncertainty relations can be reformulated in the stochastic variational method. In this approach, the finite minimum value of the uncertainty of position and momentum is attributed to 
the non-differentiable (virtual) trajectory of a quantum particle, and both of the Kennard and Robertson-Schr\"{o}dinger inequalities in quantum mechanics are reproduced \cite{koide18}.
The similar non-differentiable trajectory is considered for the trajectory of fluid elements in hydrodynamics. 
By applying the same procedure to the position and the momentum of fluid elements, 
the Kennard-type and Robertson-Schr\"{o}dinger-type uncertainty relations are obtained for the fluid described by the Navier-Stokes-Fourier equation \cite{koide18}. 
These relations are applicable to the trapped Bose gas described by the Gross-Pitaevskii equation and 
then the field-theoretical uncertainty relation is reproduced.

The derived Kennard-type inequality was numerically investigated with the initial condition of the static Gaussian mass distribution. 
We consider two different cases: one is $\xi > \nu$ and the other $\xi < \nu$, where $\xi$ is the kinematic viscosity and $\nu$ is identified with the diffusion coefficient.
The former case will correspond to liquid and the latter to gas.
We found that the behaviors of the uncertainty relations for liquid and gas are qualitatively different: 
the products of the standard deviations for liquid are always larger than that for the ideal fluid while those for gas are always smaller.

These numerical results suggest that the difference of liquid and gas is characterized through 
the behavior of the uncertainty relations.  
The fluid with a larger uncertainty than that of the ideal fluid behaves as liquid, while the one with a smaller uncertainty behaves as gas.
For a given $\alpha_A = \xi/\nu$, a critical Reynolds number $\RE^*$ may be defined. The fluid of $\RE > \RE^*$ behaves as liquid while that of $\RE < \RE^*$ is gas.
For the cases of $\alpha_A =0.1$ and $10$, $\RE^*$ will be $\infty$ and $0$, respectively.
A finite number of $\RE^*$ will be observed for the fluid of $\alpha_A$ close to  $1$.

To establish this conjecture, we have to investigate the behaviors in more general initial conditions, equation of states and higher spatial dimensional systems. Moreover, 
our definition of the standard deviation depends on the choice of fluid elements when initial conditions are asymmetric.
This does not stand in the way of finding the generalized relations in hydrodynamics, but, 
to improve the above classification, 
it is worth considering another definition which is independent of the choice of fluid elements for any initial condition.

There are still many properties of the uncertainty relations which should be studied.
The minimum uncertainty state in hydrodynamics is one of examples.
In quantum mechanics, the coherent (squeezed) state is known as the minimum uncertainty state but 
the corresponding state in hydrodynamics is not known \cite{gazeau_book}.

We have considered the standard deviations of position and momentum which are averaged over all fluid elements. 
On the other hand, the uncertainty for a single fluid element will be important in other applications.
For example, the particle production from an excited vacuum is studied in relativistic heavy-ion collisions. 
In principle, the dynamics of these particles is described by quantum chromodynamics (QCD), but the non-equilibrium behavior of QCD is very difficult to describe. Thus, as an effective approach, the excited vacuum is approximately replaced by a classical continuum medium which is described by relativistic hydrodynamics. This effective model describes the distributions of the produced particles successfully \cite{kodama-review}. 
In extracting the information of particles from the classical fluid, we assume a hadronization mechanism where 
the particles are produced by thermal radiation from each fluid element which is thermally equilibrated.
In this mechanism, the motion of the fluid element is simply assumed to coincide with the streamline of the relativistic fluid, but 
it is unlikely for viscous fluids as was shown in Fig.\ \ref{fig:bm_fluidelement}.
The uncertainty for a single fluid element will be used to compensate this discrepancy. 
Moreover the uncertainty relations are associated with microscopic behaviors which are coarse-grained in hydrodynamics and hence 
may provide more detailed information of the statistical distribution of quarks and gluons which are the constituent particles of the relativistic fluid beyond the standard hydrodynamical analysis.

We have so far considered fluids described in Cartesian coordinates.
The uncertainty relations in generalized coordinates is known to be difficult. 
For example, the uncertainty relation for angle has a famous paradox in quantum mechanics \cite{carr_review}.
Let us consider a 2-dimensional system described in polar coordinates. Then the conjugate momentum associated with the angular variable is given by the angular momentum.  
If one defines an angle operator $\theta_{op}$ such that its commutation rule with the angular momentum operator $L_{op}$ is canonical, $[\theta_{op},L_{op}] = i\hbar$, then the obtained Kennard inequality reads 
\begin{eqnarray}
\sigma_\theta  \sigma_L  \ge \frac{\hbar}{2}\, ,
\end{eqnarray}
where $ \sigma_\theta$ and $ \sigma_L $ are defined by Eq.\ (\ref{eqn:qm_sd}) with the Hilbert vector .  
This is however unacceptable because $\sigma_\theta$ should be infinite for the eigenstate of $L$ but the maximum value of $\sigma_\theta$ is finite due to the bounded domain of the spectrum, $0\le \theta < 2\pi$.

This discrepancy is usually attributed to the difficulty of the introduction of a multiplicative angle operator satisfying the standard canonical commutation rule. See a review paper \cite{carr_review} and the recent papers \cite{ohnuki, tanimura,kas06,jp1,diego18} for survey and discussion. 
On the other hand, the procedure developed in this paper
need not to introduce the operator representations of position and momentum and thus we can avoid the difficulty associated with the definition of operators.  
In Ref.\ \cite{koide20-1}, the uncertainty relation in generalized coordinates 
is obtained. 
For a contravariant position vector $q^i$ and a covariant momentum vector $p_j$ in a D-dimensional system, 
the corresponding Kennard inequality is given by 
\begin{eqnarray}
\left( \sigma^{(2)}_{q^{i}} \right)^{1/2} \left( \sigma^{(2)}_{p_{j}} \right)^{1/2}
& \ge&
\frac{\hbar}{2}
\sqrt{
\left| 
\delta^{i}_j
- \int \ud^D q\, \partial_j \{ J \rho \left(q^{i} - E\left[\widehat{q}^{\,i}_t\right]\right)\}
+  \int J \ud^D q\, \rho\,  \left( q^{i} - E\left[\widehat{q}^{\,i}_t \right] \right) \Gamma^k_{jk} 
 \right|^2} \, , \label{eqn:ucr-gene} \nonumber \\
\end{eqnarray}
where we used Einstein's notation of the summation, and $J$ and $\Gamma^i_{jk}$ are the Jacobian and the Christoffel symbol, respectively.
One can confirm that the paradox for the angular uncertainty relation does not exist in this inequality. 
Moreover, SVM is applicable to quantum systems in curved geometry \cite{koide19}.
The above uncertainty relation is applicable to curved geometries and hence 
its generalization to hydrodynamics may be useful to investigate the properties of, for example, 
highly dense quark-hadron matter in binary neutron star mergers 
which are described in general relativistic hydrodynamics \cite{stocker1,stocker2}.

\vspace{3cm}

T.\ Koide acknowledges the financial support by CNPq (303468/2018-1).
A part of the work was developed under the project INCT-FNA Proc.\ No.\ 464898/2014-5.

\appendix 

\section{Ito's lemma} \label{app:ito}

Let us consider an arbitrary smooth function $f({\bf x},t)$. 
Then the change of this function on a stochastic particle, which is described by the forward SDE (\ref{eqn:fsde}), is given by 
Ito's lemma, 
\begin{eqnarray}
\ud f(\widehat{\bf r}({\bf R},t),t) &= \ud  t \left( \partial_t + {\bf u}_+ (\widehat{\bf r}({\bf R},t),t)\cdot \nabla + \nu \nabla^2 \right)f(\widehat{\bf r}({\bf R},t),t)  \nonumber \\
&+ \sqrt{2\nu} \nabla f(\widehat{\bf r}({\bf R},t),t) \cdot \ud \widehat{\bf W}(t) \, + o(\ud t) \, .
\end{eqnarray}
This corresponds to the lower series truncation of the Taylor expansion in the stochastic calculus.
Note that the Wiener process has a dimension $\ud \widehat{\bf W}(t) \sim \sqrt{\ud t}$, 
and the above result has a part of the contribution  of the second order in $\ud \widehat{\bf r}({\bf R},t)$.

The same argument is applied when the stochastic particle is described by the backward SDE (\ref{eqn:bsde}) ,  
\begin{eqnarray}
\ud f(\widehat{\bf r}({\bf R},t),t)  &= \ud  t \left( \partial_t + {\bf u}_- (\widehat{\bf r}({\bf R},t),t)\cdot \nabla - \nu \nabla^2 \right)f(\widehat{\bf r}({\bf R},t),t) \nonumber \\
&+ \sqrt{2\nu} \nabla f(\widehat{\bf r}({\bf R},t),t) \cdot \ud \underline{\widehat{\bf W}}(t) \, + o(\ud t) \, . 
\end{eqnarray}

\section{Bernoulli equation in quantum hydrodynamics} \label{app:ber}

Interestingly, the energy eigenvalue problem in the Schr\"{o}dinger equation 
corresponds to the Bernoulli equation in quantum hydrodynamics.
We find that Eq.\ (\ref{eqn:vari_qh}) can be reexpressed as
\begin{eqnarray}
\frac{1}{2}\cdot \partial_t {\bf v}^2 ({\bf x},t) = 
- {\bf v}({\bf x},t) \cdot \nabla B_{er} ({\bf x},t) \, ,
\end{eqnarray} 
where the Bernoulli function is defined by 
\begin{eqnarray}
B_{er} ({\bf x},t) = \frac{{\bf v}^2({\bf x},t)}{2} + \frac{V({\bf x})}{\uM} - 2\nu^2 \frac{\nabla^2 \sqrt{\rho({\bf x},t)}}{\sqrt{\rho({\bf x},t)}} \, .
\end{eqnarray}
For the stationary solution, the left-hand side of the above equation vanishes. 
On the other hand, ${\bf v}({\bf x},t) \cdot \nabla$ represents the differential along the flow of the probability distribution. 
Thus the Bernoulli function is constant along a streamline of the ``quantum fluid",
\begin{eqnarray}
B_{er} ({\bf x}) = \frac{{\bf v}^2({\bf x})}{2} + \frac{V({\bf x})}{\uM} - 2\nu^2 \frac{\nabla^2 \sqrt{\rho({\bf x})}}{\sqrt{\rho({\bf x})}} 
= \frac{E}{\uM} \, ,
\end{eqnarray}
where $E$ is a constant. 
On the other hand, the equation of continuity for the probability distribution (\ref{eqn:eoc}) becomes
\begin{eqnarray}
\nabla \cdot \{ \rho ({\bf x}) {\bf v} ({\bf x}) \} = 0 \, .
\end{eqnarray}
Using the definitions of the phase (\ref{eqn:phase}) and the wave function (\ref{eqn:wf}), 
 the above two equations lead to 
\begin{eqnarray}
\left(
- 2\uM \nu^2 \nabla^2 + V({\bf x})
\right) \Psi ({\bf x}) = E \Psi ({\bf x}) \, .
\end{eqnarray}
One can easily see that this is the time-independent Schr\"{o}dinger equation 
by choosing $\nu = \hbar/(2\uM)$.

\section{Relation to quantum mechanical standard deviation}
\label{app:relationtoqm}

We will show that Eq.\ (\ref{delta_p}) is equivalent to that in quantum
mechanics. See also the discussion in Ref.\ \cite{falco}. First we find the
second order correlation of the momentum operator in quantum mechanics as
\begin{eqnarray}
\langle \mathbf{p}^{2}_{op} \rangle 
= \langle - i \hbar \nabla \rangle 
= \langle\hbar^{2}
(\nabla \ln \sqrt{\rho ({\bf x},t)})^{2} + (\nabla \theta (\mathbf{x},t))^{2} \rangle,
\end{eqnarray}
where $\langle~~~~\rangle$ represents the expectation value with the wave
function with the decomposition (\ref{eqn:wf}).
On the other hand, Eq.\ (\ref{eqn:cc}) is re-expressed with this decomposition of
the wave function as
$\uM(\mathbf{u}_+ (\mathbf{x},t) - \mathbf{u}_- (\mathbf{x},t)) = \hbar\nabla \ln \rho (\mathbf{x},t)$.
Note that the current of the probability distribution is equivalent to that of the
Fokker-Planck equation. Therefore the mean velocity is expressed as
$\uM  \mathbf{v} (\mathbf{x},t) 
= \nabla \theta (\mathbf{x},t)$.

Using these relations, we can show 
\begin{eqnarray}
\frac{ \lceil (\uM\widehat{\mathbf{u}}_+)^{2} + (\uM \widehat{\mathbf{u}}_-)^{2} \rfloor}{2} = \langle \mathbf{p}^{2}_{op} \rangle\, ,
\end{eqnarray}
and
\begin{eqnarray}
\lceil \uM \widehat{\mathbf{u}}_\pm \, \rfloor = \langle \mathbf{p}_{op} \rangle \, .
\end{eqnarray}
Summarizing these results, we find
\begin{eqnarray}
\langle \mathbf{p}^{2}_{op} \rangle- \langle \mathbf{p}_{op} \rangle^{2}  
= \frac{ \lceil (\uM \widehat{\bf{u}}_+)^2 \rfloor - ( \lceil \uM \widehat{\bf{u}}_+ \, \rfloor)^{2} }{2} 
+ \frac{ \lceil (\uM \widehat{\bf{u}}_-)^2 \rfloor - ( \lceil \uM \widehat{\bf{u}}_- \, \rfloor)^{2} }{2} \, .
\end{eqnarray}
The right-hand side of the
above result is equivalent to Eq.\ (\ref{delta_p}).

\section{The positivity of the kinetic term of the Lagrangian} \label{app:positivity}

To obtain the NSF equation, we chose $\alpha_{A} > 0$ and $\alpha_{B} = 0$, leading to
\begin{equation}
\mathrm{Tr} ({\cal M})>0~\mathrm{and}~\det ({\cal M})<0,
\end{equation}
where ${\cal M}$ is defined by Eq.\ (\ref{eqn:cal_m}). 
On the other hand, from the well-known second-order algebra, the necessary and
sufficient condition for the positive-semidefinite kinetic term in
the Lagrangian is given by
\begin{equation}
\mathrm{Tr} ({\cal M})>0~\mathrm{and}~\det ({\cal M})>0,
\end{equation}
and it demands
\begin{equation}
\xi <\sqrt{\kappa}. \label{mu<k}%
\end{equation}
Therefore, to have a finite viscosity requiring the positive kinetic term,
$\kappa$ should not vanish but this is not the case of the standard NSF equation. 
Note, however, that it is not clear whether such a
requirement is mandatory, because the above positivity is irrelevant to the
positivity of the fluid energy which is defined by 
\begin{eqnarray}
\int d^{3} \mathbf{x} \left[  \frac{1}{2} n (\mathbf{x},t) \mathbf{v}^{2} (\mathbf{x},t) + \varepsilon(n(\mathbf{x},t)) \right] \, ,
\end{eqnarray}
and positive independently of the value of $\kappa$.

\section{Gross-Pitaevskii equation} \label{app:gp}

As we discussed, the quantization of a classical system can be regarded as the stochastic optimization 
of the corresponding classical action.
Then it is interesting to investigate the quantization of the ideal fluid 
where $(\alpha_A,\alpha_B,\nu) = (0,1/2,\hbar/(2\uM))$.
We further set
\begin{eqnarray}
\varepsilon(n({\bf x},t)) = \frac{U_0}{2} \frac{n^2({\bf x},t)}{\uM^2} \, , \label{eqn:gp-ener}
\end{eqnarray}
where $U_0$ is the coupling constant of the two-body interaction. 
Substituting these into Eq.\ (\ref{eqn:vari-con-med1}) and 
introducing the wave function for the continuum medium as 
\begin{eqnarray}
\Psi ({\bf x},t) = \sqrt{\frac{n ({\bf x},t)}{\uM}} e^{\ii \theta ({\bf x},t)} \, 
\end{eqnarray}
we find 
\begin{eqnarray}
\ii \hbar \partial_t \Psi ({\bf x},t)  = \left[
-\frac{\hbar^2}{2\uM} \nabla^2+ U_0 |\Psi ({\bf x},t)|^2
\right] \Psi ({\bf x},t) \, . \label{eqn:gpequation}
\end{eqnarray}
This equation is known as the Gross-Pitaevskii equation. 
In this derivation, we need not to assume the local thermal equilibrium 
and thus SVM is applicable directly to the 
trajectory of a constituent particle, instead of the fluid element. 
Then $\uM$ is given by the mass of the constituent particle.

One may wonder about the difference between $\varepsilon$ and the second term 
on the right-hand side of Eq.\ (\ref{eqn:gpequation}).
This is due to the difference between the left-hand sides of Eqs.\ 
(\ref{eqn:vari-particle1}) and (\ref{eqn:vari-con-med1}), 
\begin{eqnarray}
\frac{1}{\uM} \nabla V ({\bf x},t)\longleftrightarrow \frac{1}{n({\bf x},t)} \nabla P(n({\bf x},t)) \, .
\end{eqnarray} 
If the spatial dependence of $n({\bf x},t)$ is small, we approximately find 
\begin{eqnarray}
\frac{1}{n({\bf x},t)} \nabla P(n({\bf x},t)) \approx  \nabla \frac{P(n({\bf x},t))}{n({\bf x},t)} \, .
\end{eqnarray}
Note that the similar procedure is well-known in the derivation of the Bernoulli equation of classical fluids.
Then the above Gross-Pitaevskii equation is replaced with 
\begin{eqnarray}
\ii \hbar \partial_t \Psi ({\bf x},t)  = \left[
-\frac{\hbar^2}{2\uM} \nabla^2+ \frac{U_0}{2} |\Psi ({\bf x},t)|^2
\right] \Psi ({\bf x},t) \, .
\end{eqnarray}
Here the potential term is the same as Eq.\ (\ref{eqn:gp-ener}).
The above discussion will be related to the different behavior of the stochastic Hamiltonian 
in Sec.\ \ref{sec:hami} and the Bernoulli equation in Appendix \ref{app:ber}. 

It is worth mentioning that 
the application of SVM to the derivation of the Gross-Pitaevskii equation in a many-particle system is studied in Ref.\ \cite{GP_loffredo}.

\section{Smoothed particle hydrodynamics} \label{app:sph}

We briefly summarize Smoothed Particle Hydrodynamics (SPH), which is a numerical calculation method to solve hydrodynamics \cite{monaghan,lucy}. 
To implement a numerical simulation, we have to reexpress the NSF equation.
For example, a standard procedure is to replace the differential equation with the corresponding difference equation by introducing spatial grids.
In SPH, however, hydrodynamical quantities are expressed by the ensemble of a finite number of quasi-particles and then hydrodynamics is mapped into the motions of the particles.
The introduced quasi-particle is called SPH particle.

The SPH particle has a finite volume with is characterized by a length parameter $h$. 
Therefore the hydrodynamical behaviors which are smaller than this scale $h$ are coarse-grained.
Let us consider the SPH representation of the mass distribution.
To represent a fluid with $N_{SPH}$ particles, each SPH particle should have a mass $\chi$
which is defined by 
\begin{eqnarray}
\chi = \frac{\uM_T}{ N_{SPH} } \, ,
\end{eqnarray}
where $\uM_T$ is the total mass of the fluid.

\begin{figure}[t]
\includegraphics[scale=0.25]{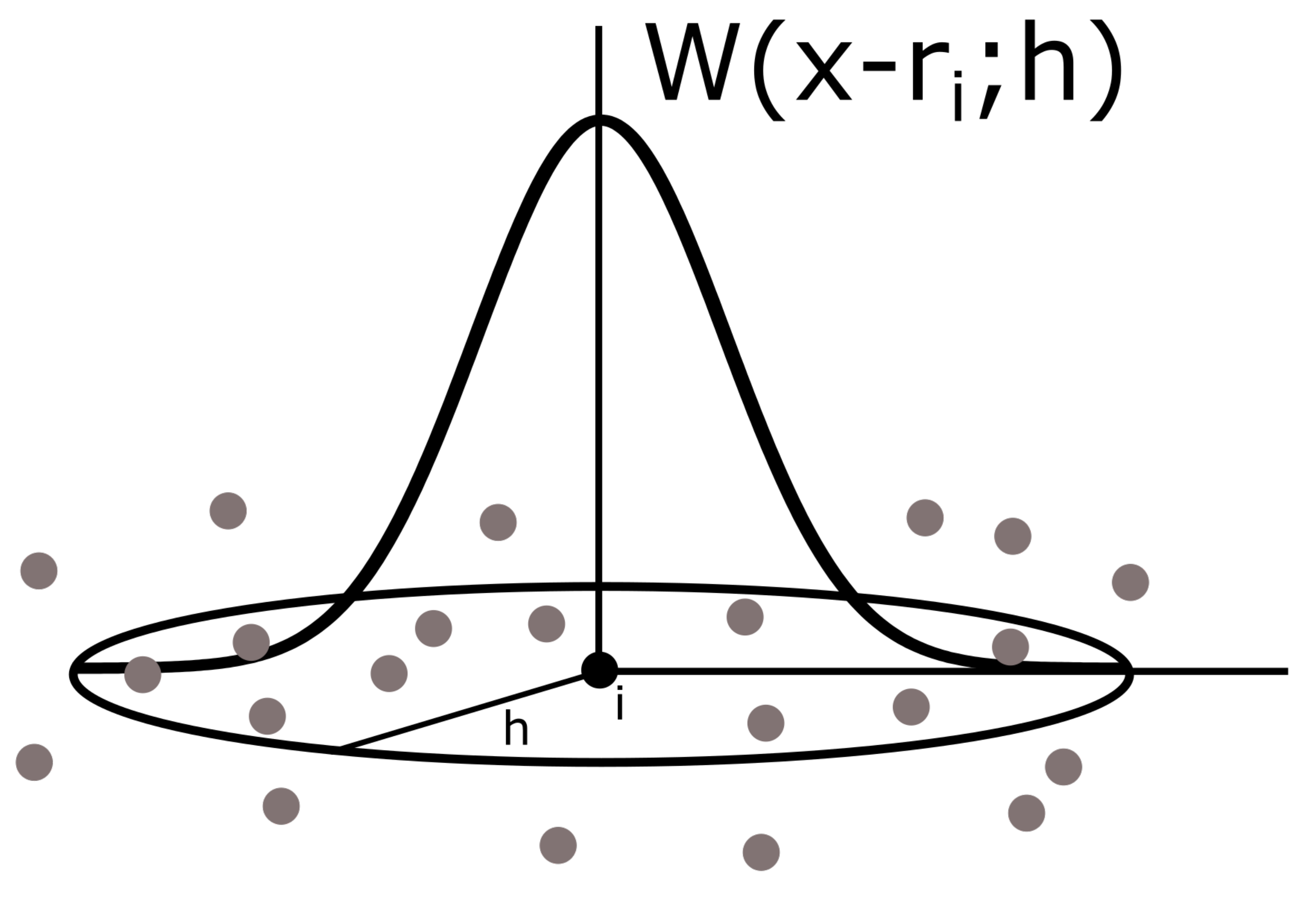}
\caption{The $i$-th SPH particle has an overlap with the $j$-th particle located in the region of $|{\bf x}_j - {\bf x}_i| \le h$.}
\label{fig:kernel}
\end{figure}

Then the coarse-grained mass distribution is given by the sum of the contributions from all SPH particles,
\begin{eqnarray}
n ({\bf x},t) = \sum_{i=1}^{N_{SPH}} \chi W ({\bf x}-{\bf r}_i (t);h) \, , \label{eqn:sph_mass}
\end{eqnarray} 
where 
${\bf r}_i (t)$ denotes the trajectory of the $i$-th SPH particle and 
the smoothing function $W ({\bf x};h)$ satisfies the properties,
\begin{equation}
\begin{split}
& W({\bf x};h) = 0 \, \, \, \,  (|{\bf x}| > h) \, ,\\
& \lim_{h \rightarrow 0}W({\bf x};h) = \delta^{(D)} ({\bf x}) \, ,\\
& \int \ud^D x \, W ({\bf x};h) = 1\, .
\end{split}
\end{equation}
This function characterizes the volume of the SPH particle 
and plays a role of a form factor.
Thus the microscopic oscillations of the mass distribution which are smaller than $h$ are smoothed out in SPH.
See also Fig. \ref{fig:kernel}.
There are various candidates of the smoothing function. For example the quintic spline function satisfies 
the above requirements,  
\begin{eqnarray}
W({\bf x};h) 
= 
N \left\{
\begin{array}{ll}
(3-q)^5 - 6 (2-q)^5 + 15 (1-q)^5 & 0 \le q < 1/3 \\
(3-q)^5 - 6 (2-q)^5 & 1/3 \le q < 2/3 \\
(3-q)^5 & 2/3 \le q \le 1 \\
0 & 1 < q \, ,
\end{array}
\right.
\end{eqnarray}
where $q = |{\bf x}|/h$ and the normalization factor $N$ is $1/(40h)$, $63/(478 \pi h^2)$ and $81/(359 \pi h^3)$ 
for one, two and three dimensional systems, respectively.
For the simulations in this paper, we choose $h/x_0 = 0.24$ and $N_{SPH}=4000$.

One can see that the time dependence of the mass distribution (\ref{eqn:sph_mass}) is represented through that of the SPH-particle trajectory. 
As a matter of fact, the equation of the trajectory is determined from the NSF equation. 
The SPH representation of the NSF equation in 1+1 dimension is given by 
\begin{eqnarray}
\frac{\ud^2 r_{i}}{\ud t^2} &=&
-\sum_{j=1}^{N_{SPH}} \chi \left(\frac{\Pi_{i}}{n_{i}^{2}(r_i) }+\frac{\Pi_{j}}{n_{j}^{2} (r_j) }\right)\partial_{r_i}W (r_i -r_j;h ) \, ,
\end{eqnarray}
where 
\begin{eqnarray}
\Pi_i &=& P(n(r_i)) - \frac{\eta(r_i)}{n(r_i)} \sum_{j=1}^{N_{SPH}} \chi \left( \frac{\ud r_j}{\ud t} - \frac{\ud r_i}{\ud t}  \right) 
\partial_{r_i} W (r_i -r_j;h ) \, .
\end{eqnarray}
One can see that the above equation does not have the nonlinear velocity term associated with ${\bf v}\cdot \nabla {\bf v}$ in the NSF equation. 
This is because this term is absorbed into the material derivative $\ud/\ud t$ of the above equation. 
This is one of the advantages of  the SPH scheme. Another advantage is that the equation of continuity of the fluid mass is automatically satisfied by Eq.\ (\ref{eqn:sph_mass}).

\end{document}